*Under consideration for publication in J. Fluid Mech.*

# Vortex promoters in magnetohydrodynamic duct flows

**Andreu Queralt, Dmitry Krasnov, Yuri Kolesnikov and Jörg Schumacher**
Institute of Thermodynamics and Fluid Mechanics, Technische Universität Ilmenau, P.O.Box 100565, 98684 Ilmenau, Germany
**Corresponding author:** Andreu Queralt, andreu.queralt-mcbride@tu-ilmenau.de

Liquid metals are a prime candidate for cooling the first wall of future fusion reactors, realized by cooling blankets. These are arrays of long rectangular ducts in different orientation subject to the strong magnetic field that holds the fusion plasma. It is well known that strong magnetic fields suppress mixing, as the critical Reynolds number increases and vortices are stretched along the magnetic field lines, which we consider transverse to the mean flow. This results in quasi-two-dimensional (Q2D) states or even complete vortex decay. Here, we study the effects of vortex promoters in blankets to generate these Q2D states by means of three-dimensional direct numerical simulations of quasistatic magnetohydrodynamics with both, electrically insulating and conducting walls. In case of insulating walls, disturbances are found to be sustained. When the duct walls are conducting, the disturbances may be extinguished immediately and replaced by a Walker- or Hunt-type flow, which develops instabilities at sufficiently large Hartmann numbers and given the duct is long enough. Furthermore, we add heat transfer to the simulations by imposing constant heat flux at the Shercliff side walls. We classify and compare possible configurations with respect to their turbulent transport properties. For highly conducting ducts, no additional effect is observed. For insulating ducts in horizontal position, we observe a small effect. For vertically oriented ducts, the buoyancy forces have a significant impact by adding new buoyancy driven instabilities to the flow, which produce intermittent fluctuations. We analyse the turbulent kinetic energy (TKE) and the Nusselt number $Nu$ of each of these flows and find those with the largest TKE may not have the best heat transfer performance. This is caused by side jets that remove heat quickly without necessarily mixing with the bulk flow. The buoyancy force and wall conductance ratio are found to play a key role in determining the flow structure. Part of our work is a parametric study at fixed Reynolds, Hartmann and Prandtl numbers, which allows us eventually to compose a phase diagram showcasing the different flow regimes.

**Key words:** High-Hartmann-number flows, Turbulence simulation, Vortex instability

## 1. Introduction

Liquid metal (LM) blankets serve a dual purpose in the context of fusion reactors. First, they are used to cool the extremely hot surfaces of the first wall of the vacuum chamber, where the plasma conditions are adequate for the fusion reactions to occur (Abdou *et al.* 2015), and secondly, the liquid metal can breed tritium, part of the fuel needed for the reaction to occur (Giancarli *et al.* 2020). This is accomplished due to the neutrons from the nuclear fusion reaction, which are not confined to the chamber by the magnetic field and can therefore penetrate the wall and enter the liquid metal where heat is transferred thus allowing breeding to occur. The presence of neutrons at a certain distance from the wall decays exponentially (Smolentsev *et al.* 2008; Molokov *et al.* 2007).

The coolant is pumped into ducts arranged along the walls of the toroidal chamber (Tokamak) where the plasma is heated. To confine the plasma to the fusion chamber, a three-dimensional magnetic field is required. To a good approximation, the magnetic field can be considered one-dimensional and uniform for the blanket segments, which is what we will apply for our study. The blankets are basically rectangular ducts, which are arranged in either the poloidal or toroidal field directions (Giancarli *et al.* 2020). We will assume here, that the ducts are arranged along the poloidal magnetic field direction. Finally, different duct geometries have been proposed, some consist of curved segments to conform to the reactor walls, such as in the so-called "banana" configuration (Smolentsev 2021). In the following we will consider straight duct segments only (Molokov *et al.* 2007; Smolentsev *et al.* 2008).

From an energetic point of view, the pressure drop $\Delta p$ in the entire LM circuit has to be minimized. Even though most of the pressure drop is caused by sharp corners (due to expansions, bent overs, see Mistrangelo *et al.* (2021); Mistrangelo & Bühler (2024)) when entering and exiting straight segments, efforts to reduce the pressure drop even in the ducts have led researchers to include a flow channel insert (FCI) made of ceramic materials at the duct walls, to make them electrically insulating, see, e.g., Smolentsev (2021). However, the concept of FCI also poses a number of problems. Chief among them is that the ceramic material is destroyed by the incoming neutron flux at a high rate, which would increase the maintenance requirements for the power plant (Smolentsev *et al.* 2008; Molokov *et al.* 2007). This makes a study of both, insulating and conducting ducts, relevant.

High velocity gradients at the first wall have also been linked to increased corrosion (Smolentsev *et al.* 2013, 2024) and to a reduction of the amount of breeded tritium (Rubel 2019). This is especially challenging for cases with electrically conducting walls, where the Walker flow forms near wall jets with strong gradients.

To enhance heat transfer, the use of vortex promoters has been proposed and studied both, numerically and experimentally, see e.g. Hamid *et al.* (2016) and Krasnov *et al.* (2023*b*). This idea has been broadly addressed for various flow settings, including isolated boundary layers (Yuan *et al.* 2025) and duct flows (Zhang *et al.* 2024; Bao *et al.* 2025). Several attempts of flow control and optimization have been undertaken, e.g. by He *et al.* (2023) and Wang *et al.* (2025). A likely candidate to be used as a vortex promoter is a cylinder parallel to the magnetic field. The disturbances introduced to the flow by these vortex promoters are akin to the phenomenon of *magnetoconvective fluctuations* (MCF), see Genin *et al.* (2011), which appear in a wide range of flow parameters and configurations, before strong magnetic fields can suppress them. MCFs are characterized by high-amplitude and low-frequency oscillations in the velocity and temperature fields. They appear randomly and are quite pertinent to systems that combine thermal convection and magnetic fields, as shown in experimental and numerical studies (Sviridov *et al.* 2010; Zikanov *et al.* 2013; Zhang & Zikanov 2014; Mistrangelo & Bühler 2014). Even though MCFs can mix the flow and enhance heat transport, their random nature and extreme gradients are detrimental not only to the operation of the exchanger, but also to its mechanical structure. The use of vortex promoters is, therefore, intended both to extend the parameter space in which these structures evolve and to impart a predictable and regular character to MCFs (Belyaev *et al.* 2023), without hazardous random bursts. This sets the motivation for the present study.

Here, we study the viability of these vortex promoters by performing three-dimensional direct numerical simulations (DNS) of quasistatic magnetohydrodynamics (MHD) in liquid metal flows in rectangular ducts at different Reynolds and Hartmann numbers. To this end, we consider differently oriented ducts in the presence of a strong transverse magnetic field with and without additional buoyancy effects. The walls of the duct are either electrically insulated or conducting. We classify the different flow regimes and evaluate low-order turbulence statistics, such as mean velocity profiles, fluctuations, and Reynolds stresses. The balance between buoyancy and Lorenz forces is found to determine the flow structure. Furthermore, the electrical conductivity of the wall rather than the strength of the magnetic field is essential. Our DNS show that the distribution of the Lorenz force becomes highly relevant when both effects are of comparable magnitude. Furthermore, we analyse the turbulent momentum and heat transfer for the given configurations. Finally, we build a phase diagram which summarizes the transition between flows with varying Grashof number $Gr$ and the wall conductance ratio of the duct $c_W$ (both defined in the next section).

The outline of the manuscript is as follows. In section 2, we introduce the equations of motion for the quasistatic MHD case, the parameter space, and the different duct configurations. In section 3, we present the results of DNS conducted at the extreme ends of the parameter range. In section 4, we compare the integral turbulent kinetic energy (TKE) and the Nusselt number $Nu$ for the different runs and discuss the transitions between each of the flows from section 3 as $Gr$ and $c_W$ are changed. In section 5, we compare the turbulent kinetic energy production $P$ of each of the flows. The quasi-two-dimensional (Q2D) structures are found to have negative production $P$, indicating that they are sustained by a reverse energy cascade. When Q2D structures are not present, we see both, positive and negative production $P$. We conclude the work with a summary and an outlook to future work in the final section.

## 2. Simulation

### 2.1. *Equations of motion*

The flow is modelled using the dimensionless form of the magnetohydrodynamic (MHD) equations, in the limit of low magnetic Reynolds number ($R_m$), where the induced magnetic field $\boldsymbol{b}$ and the corresponding induction equations are neglected. This limit, known as the quasi-static MHD approximation, is widely used for liquid metal flows, see e.g., Davidson, P. A. (2001). Under these assumptions, the mass, momentum, and energy equations for an incompressible Newtonian fluid are:

$$\nabla \cdot \boldsymbol{u} = 0 \tag{2.1}$$

$$\frac{\partial \boldsymbol{u}}{\partial t} + (\boldsymbol{u} \cdot \nabla)\boldsymbol{u} = -\nabla p + \frac{1}{Re}\nabla^2 \boldsymbol{u} + \frac{Ha^2}{Re}(\boldsymbol{j} \times \boldsymbol{e_B}) - \frac{Gr}{Re^2} T \boldsymbol{e_g} \tag{2.2}$$

$$\frac{\partial T}{\partial t} + (\boldsymbol{u} \cdot \nabla)T = \frac{1}{RePr}\nabla^2 T \tag{2.3}$$

Here $\boldsymbol{u}$, $p$, $\boldsymbol{j}$, and $T$ are the velocity, pressure, electric current density and temperature of the fluid, $\boldsymbol{e_B}$ and $\boldsymbol{e_g}$ are the unit vectors in the direction of the magnetic field and gravity.

The system is closed with Ohm's law for electric current density $\boldsymbol{j}$ and electric potential $\phi$, determined by charge conservation $\nabla \cdot \boldsymbol{j} = 0$

$$\boldsymbol{j} = -\nabla\phi + \boldsymbol{u} \times \boldsymbol{e_B} \tag{2.4}$$

$$\nabla^2\phi = \nabla \cdot (\boldsymbol{u} \times \boldsymbol{e_B}). \tag{2.5}$$

The parameters $Re$, $Ha$, $Gr$ and $Pr$ in (2.1) – (2.5) are the dimensionless Reynolds, Hartmann, Grashof and Prandtl numbers (see Table 1 for definitions). An important dimensionless parameter here is the Hartmann number $Ha$, which estimates the ratio between the Lorentz and viscous forces. Another MHD parameter is the Stuart number (or interaction parameter) $N = Ha^2Re^{-1}$, which estimates the ratio between the Lorentz and inertia forces. This study aims at systems with $Ha \gg 1$ and $N > 1$.

Finally, boundary conditions for velocity, temperature, and electric potential must be prescribed. For velocity, we apply the no-slip boundary condition $\boldsymbol{u} = 0$ at the walls, and the convective outflow condition $\partial\boldsymbol{u}/\partial t + U_0\partial\boldsymbol{u}/\partial x = 0$ at the exit, here $U_0$ is the flux-based mean velocity. For temperature conditions, the sidewalls have a prescribed constant heat flux $q_w$, and the Hartmann walls are assumed adiabatic $q_w = 0$

$$q_w = \lambda \left. \frac{\partial T}{\partial n} \right|_w . \tag{2.6}$$

At the inlet the temperature is specified as an unperturbed uniform profile $T = T_0$ ($T_0 = 0$ in our non-dimensional units) and at the exit we apply the convective outflow condition $\partial T/\partial t + U_0\partial T/\partial x = 0$, as for the velocity field, see Belyaev *et al.* (2020) for more details on outflow conditions.

The electric potential boundary condition is modeled via the thin wall approximation (Sterl 1990), given by:

$$\left. \frac{\partial \phi}{\partial n} \right|_w = c_W \nabla_\perp^2 \phi. \tag{2.7}$$

This condition holds when the wall thickness $\tau_w$ is much smaller than the thickness of the fluid layer $a$, i.e. $\tau_w \ll a$, and it allows us to avoid solving the electrical current density and the electric potential within the walls. The parameter $c_W$ is known as the wall conductance ratio:

$$c_W = \frac{\sigma_w \tau_w}{\sigma a}, \tag{2.8}$$

where $\sigma_w$ and $\sigma$ are the electric conductivity of the wall material and the liquid metal. The limits of $c_W = 0$ and $c_W \to \infty$ correspond to the cases of perfectly insulating and perfectly conducting walls. Usually, the conductance ratio $c_W < 1$ in experiments and in real LM blanket systems (Müller & Bühler 2001). Further details and discussions on the thin wall approximation, conductance ratio $c_W$ and its expected values can be found in (Sterl 1990; Arlt 2018; Krasnov *et al.* 2023*a*).

### 2.2. *Numerical method*

The governing equations are solved numerically with our in-house finite-difference DNS solver for incompressible flows in rectangular geometries (Krasnov *et al.* 2023*a*). The solver is based on conservative scheme (Morinishi *et al.* 1998; Ni *et al.* 2007), the spatial discretization is of $2^{nd}$ order and is performed on a structured non-uniform collocated grid. The time discretization of the momentum and heat transport equations is based on the Adams-Bashforth/backward difference of the $2^{nd}$ order. In the heat equation, the diffusion term is integrated implicitly to avoid the time step reduction due to low Prandtl numbers of $Pr \sim 10^{-2}$ typical for liquid metals. The classical projection algorithm is used to satisfy incompressibility. The computational grid is orthogonal and can be clustered in the $y$ and $z$ directions. Elliptic solvers for pressure, electric potential, and temperature use the method of separation of variables, implemented as a tensor-product approach, where full 3D problem is replaced by a series of 1D problems. This decomposition also allows to address problems with finite wall conductivities (2.7). The numerical method has demonstrated its capabilities in simulations of MHD flows at high $Ha$ and $Re$ (Zikanov *et al.* 2019; Krasnov *et al.* 2021; Listratov *et al.* 2025) and flows with heat transfer (Belyaev *et al.* 2020; Pandey *et al.* 2022). For our numerical experiments, we consider a rectangular duct of aspect ratio 3.5 : 1 and streamwise length $L_x = 32\pi$. This geometry is the same used in experiments by Belyaev *et al.* (2023). Figure 1 shows the duct geometry, here $x$ is the streamwise direction, $y$ the spanwise direction and $z$ the vertical direction of the duct. The origin is located at the center of the $y - z$ plane and at $x = 0$.

The numerical resolution used in the simulations has been set at $N_x \times N_y \times N_z = 4096\times320\times92$ points in, correspondingly, the streamwise, spanwise and vertical directions. The computational grid is uniform in the streamwise $x$-direction and non-uniform in both wall-normal $y$- and $z$-directions. Two coordinate transformations, providing grid-clustering towards the walls, have been applied, namely a modified Gauss-Lobatto (GL) stretching

$$z = \frac{L_z}{2}\left[\omega \sin\left(\frac{\pi}{2}\eta\right) + (1-\omega)\eta\right] \tag{2.9}$$

and a hyperbolic tangent transform

$$z = \frac{L_z}{2}\frac{\tanh(A\eta)}{\tanh(A)} \tag{2.10}$$

shown here for the $z$-coordinate that varies between $\pm L_z/2$. Here $-1 \le \eta \le 1$ is the uniform transformed coordinate, the weight $\omega = [0...1]$ provides linear blending between the clustered GL and uniform grids (to avoid excessive parabolic clustering at the walls), whereas the coefficient $A$ determines the degree of tanh-grid clustering. Both types of grid-stretching have been used in the prior studies of MHD flows with jets and honeycombs (Zikanov *et al.* 2019; Krasnov *et al.* 2021; Belyaev *et al.* 2020).

It should be stressed that none of the flow regimes obtained in our simulations can be classified as fully developed 3D turbulence. Instead, we obtain regimes populated with large-scale anisotropic structures, quasi-2D vortices, pseudo-laminar intermittent and transient states. Therefore, the usual requirements for the computational grid, which are dictated by the need to resolve small scale turbulent eddies, are not truly applicable here. Instead, it is crucially important to resolve the thin MHD boundary layers to provide accurate closure of the electric currents $\boldsymbol{j}$ at both the Hartmann and Shercliff walls. According to the parametric studies performed in past works (Zikanov *et al.* 2019; Belyaev *et al.* 2020; Listratov *et al.* 2025), at least 7 points per MHD boundary layer are necessary for the correct behaviour of $\boldsymbol{j}$-currents. In the present work we have used 10 points per boundary layer for the highest value $Ha = 1000$ addressed here.

In addition to that, we have also performed a grid-sensitivity study in a broad range of $Ha$ to quantify the impact of grid stretching onto flow properties, such as TKE level and integral wall shear stress $\tau_w$. Three types have been tested: (i) GL-grids in $y$- and $z$-directions with $\omega_y = \omega_z = 0.95$, (ii) tanh-grid with $A_y = 2.8$ in $y$-direction and GL-grid with $\omega_z = 0.95$ in $z$-direction and (iii) even stronger clustered tanh-grids in $y$- and $z$-directions with $A_y = A_z = 3$. The analysis of TKE, wall shear stress $\tau_w$, and also spatial decay of quasi-2D vortices revealed a visible quantitative difference between types (i) and (ii), while no differences were observed between types (ii) and (iii) (except for a much smaller integration time-step $\delta t$ in type (iii) needed to maintain numerical stability). Thus, the type-(ii) grid with tanh-clustering (2.10) at $A_y = 2.8$ between the Hartmann walls and GL-clustering (2.9) with $\omega_z = 0.95$ between the Shercliff walls has been used for the simulations presented in this paper.

| Name | Symbol | Definition | Values |
|---|---|---|---|
| Reynolds number | $Re$ | $U_0 a/\nu$ | 4000 |
| Hartmann number | $Ha$ | $B_0 a/(\sigma/\rho\nu)^{1/2}$ | 325, 500, 750 1000 |
| Grashof number | $Gr$ | $g\beta q_w a^4/\lambda\nu^2$ | 0, $10^7$ |
| Prandtl number | $Pr$ | $\nu/\lambda$ | 0.02 |

Table 1. Dimensionless parameters and their range. Here $U_0$ is the flux-based mean velocity, $a$ is the duct half-height, $\nu$ the kinematic viscosity of the fluid, $B_0$ the imposed constant magnetic field, $\sigma$ the fluid's electrical conductivity, $\rho$ the fluid density, $g$ the acceleration due to gravity, $\beta$ the isobaric expansion coefficient, $q_w$ the imposed heat flux at the walls, $\lambda$ the thermal conductivity of the fluid.

### 2.3. *Duct configurations*

We differentiate three possible duct orientations when heat transfer is present. These correspond to a horizontal duct (figure 2(a)) and to a vertical duct (figure 2(b,c)) in the latter case, the liquid metal can be pumped either upwards (figure 2(b)) or downwards (figure 2(c)). As we shall see, this has important implications for the flow properties. In all cases, the magnetic field is applied in the spanwise direction, so that $\boldsymbol{B} = (0, B_y, 0)$. The heat flux is applied to the top and bottom walls at $z = \pm 1$. In fusion reactors, the neutron flux acts as a heat source inside of the metal, and the strength of the source decays exponentially with the penetration depth Smolentsev *et al.* (2008). Experimentally, this is hard to replicate outside of fusion environments, so as an alternative heating pads are used to heat the walls. In our simulations we take the latter approach, by imposing a constant, known heat flux at both walls starting at $x = 20$, once again to have the same setup as Belyaev *et al.* (2023). While symmetric heating is unrealistic for a fusion environment, it facilitates analysis prior to studying the one-sided case. In addition, the resulting data can be used to train machine learning models to predict these type of flows at a much lower computational cost (Wang *et al.* 2025). Some of the models work best or even require symmetry (Otto & Rowley 2019). In the horizontal duct, gravity acts in the negative $z$ direction: $\boldsymbol{g} = (0, 0, -g_z)$. On the other hand, when the duct is orientated vertically, gravity will act opposite to the mean flow when it is pumped upwards ($\boldsymbol{g} = (-g_x, 0, 0)$) and in the direction of the mean flow when it is pumped downwards ($\boldsymbol{g} = (g_x, 0, 0)$). At the duct inlet ($x = 0$) we impose a velocity profile $\boldsymbol{u} = (u(y, z), 0, 0)$, where $u(y, z)$ takes the form of two flat jets that occupy roughly 50% of the duct height, as illustrated in figure 1. These jets act as vortex promoters, and a similar effect can be obtained experimentally by adding a cylinder at the inlet of the duct (Krasnov *et al.* 2023*b*).

### 2.4. *Parameter space*

Four dimensionless numbers are present in the mathematical model of section 2.1. Table 1 shows the definition of these parameters and their range in the simulations conducted. The value for $Pr = 0.02$ is typical for liquid metals (Davidson, P. A. 2001), whereas the other parameters will depend on the specifics of the reactor. For this study, the lower end of the $Ha$ number range was chosen to allow comparisons with (Krasnov *et al.* 2023*b*).

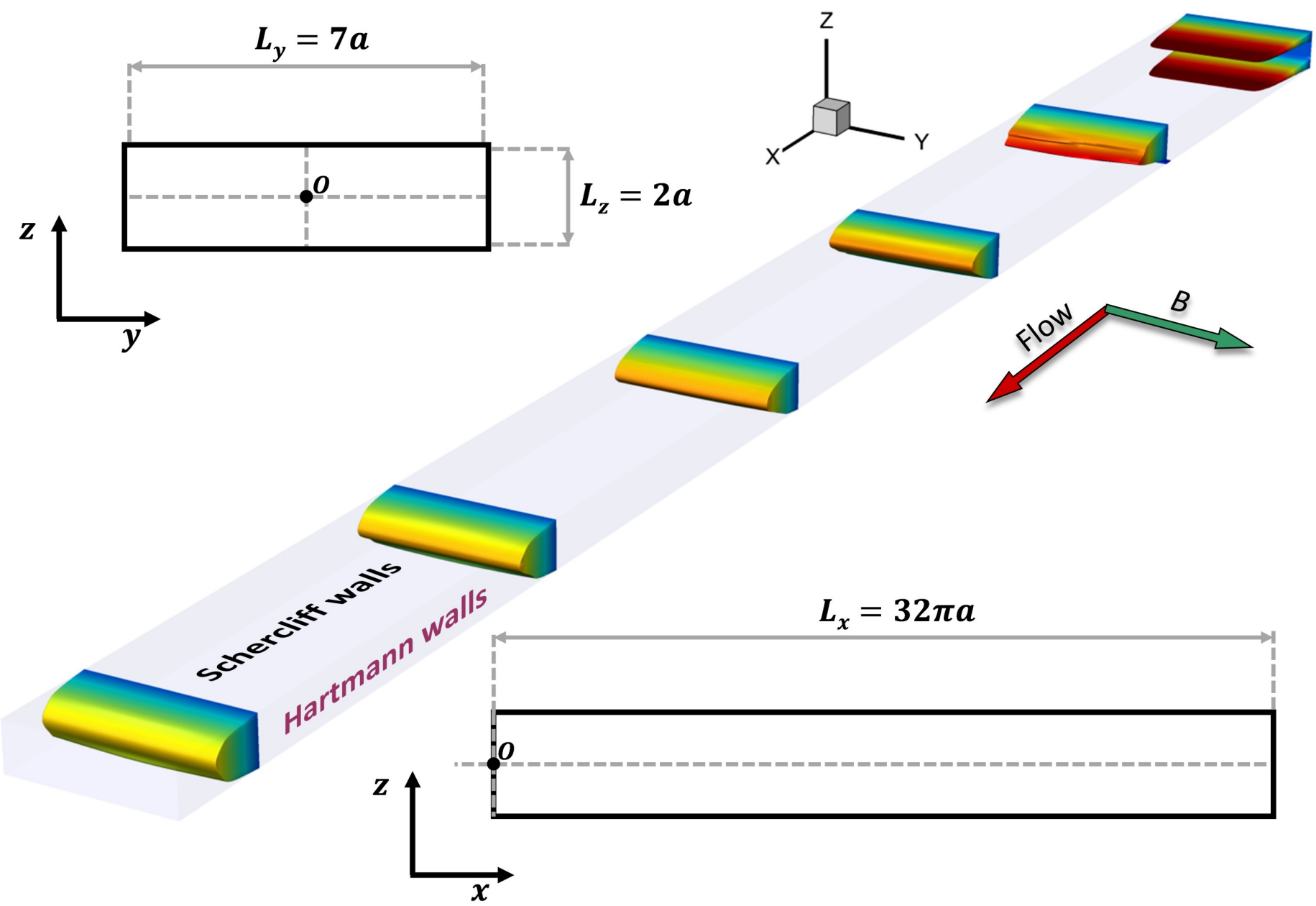

Figure 1. Principal duct geometry used in the numerical simulations. Three-dimensional view of the duct domain is shown with an example of spatial flow evolution at $Re = 4000$ and $Ha = 325$, illustrated by the velocity profiles. The profile with two flat jets prescribed at the inlet mimics flow pattern past a round cylinder. The duct dimensions and coordinate axis of the domain are also shown in the inserts at two specific cross-sections: $(y, z)$-section (top left corner) and $(x, z)$-section (low right corner). The colour arrows indicate directions of the flow and applied magnetic field.

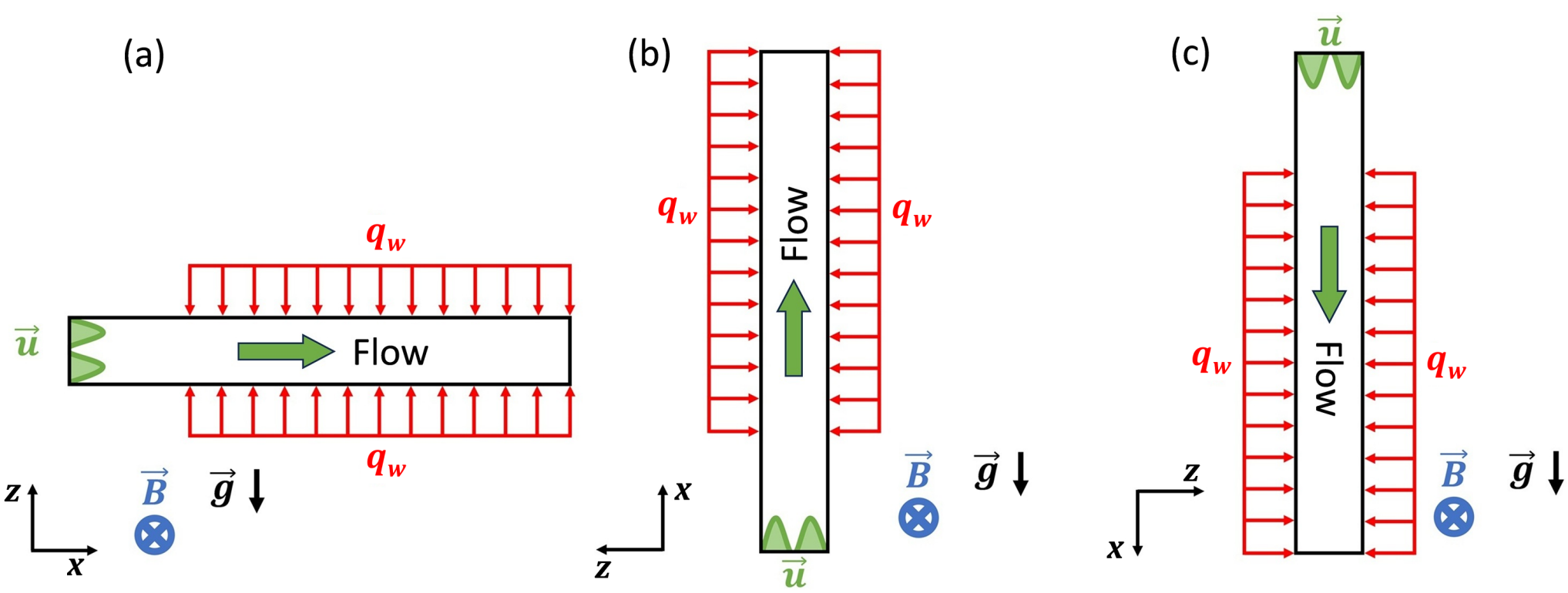

Figure 2. Duct configurations with heat transfer: horizontal (a), upwards (b) and downwards (c) flows. Note that the axis rotate with the duct.

In addition to these parameters, the boundary conditions require an electrical wall conductance ratio $c_W$. For this parameter two values are considered, $c_W = 0$ for perfectly insulating walls and $c_W = 0.1$ for highly conducting walls.

For cases where $F_b \neq 0$ the orientation of the duct also plays a role. In vertical ducts the buoyancy force $F_b$ will act in the same direction as the pressure gradient and the Lorenz force $F_L$, which will act to either accelerate or brake the flow depending on the position of the fluid parcel and the direction of the flow (upwards or downwards). For ducts in a horizontal position, the buoyancy force is normal to the pressure gradient and Lorenz force, which affects integral quantities of the flow such as TKE and $Nu$, and creates an asymmetry in the velocity profile, but overall the modifications are not as drastic as when the duct is in a vertical position, and a separate classification is not warranted.

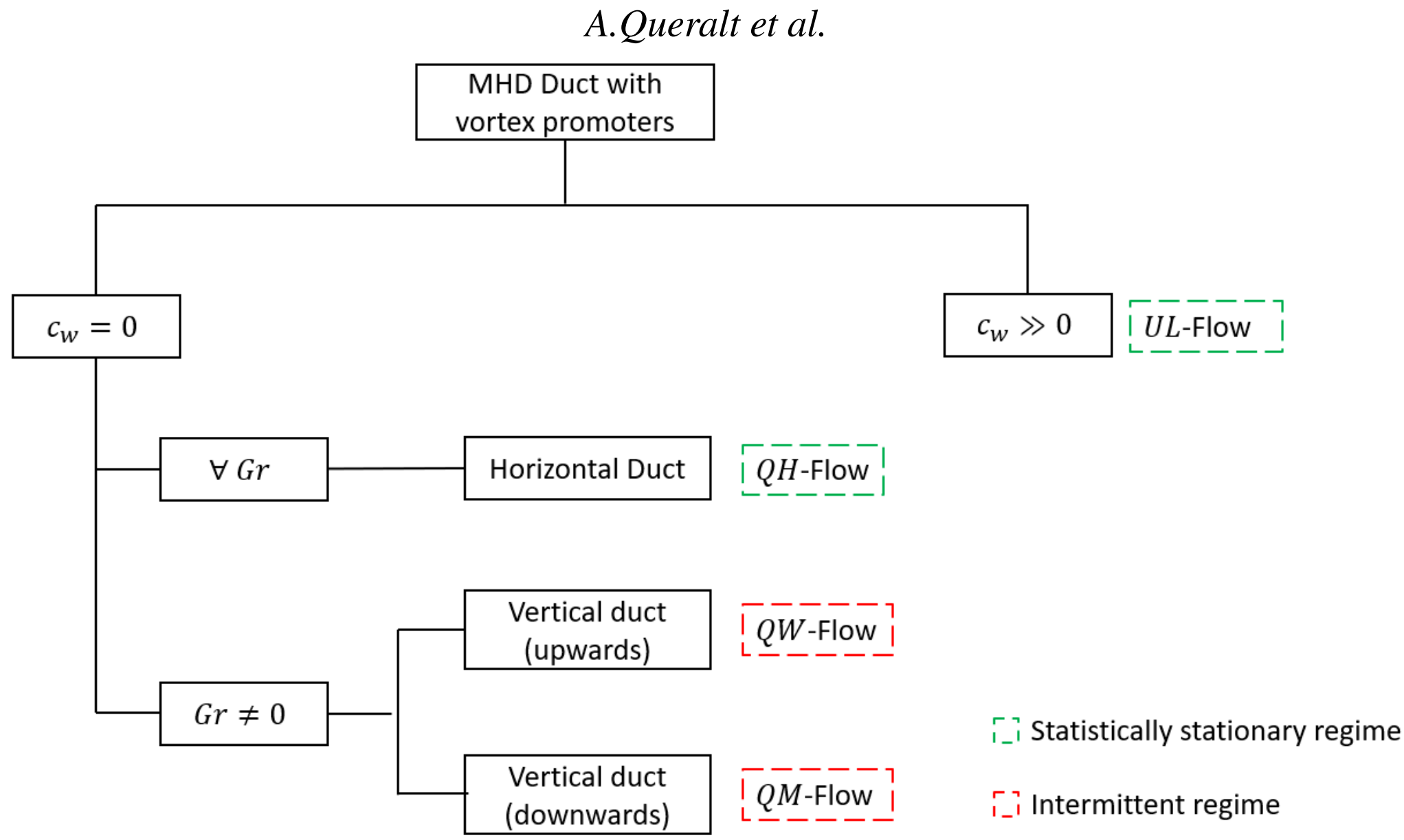

Figure 3. Schematic overview of the flow regimes in the parameter space $Re = 4000$, $Ha \in [325, 1000]$, $Pr = 0.02$, $Gr \leq 10^7$, $c_W \leq 0.1$. QH corresponds to a flow populated by Q2D structures that decay into a Hartmann flow. QW and QM correspond to a flow initially populated by Q2D structures that is replaced by side wall jets once the Q2D structures decay. W inidcates the jets are backflow jets and M that they move in the direction of the streamwise velocity. UL corresponds to a flow with unstable side jets caused by the distribution of currents in the flow.

## 3. Results

### 3.1. *Overview*

We begin by showing the results for simulations, summarized in the diagram in figure 3. The naming convention used there and throughout the paper is defined in section 3.2 and summarized in table 2 and figure 4. These simulations are conducted at the extreme ends of our parameter space defined in table 1. Four distinct types of flow are identified:

- *Quasi-2D Hartmann flow* (QH) for which the duct is populated by Q2D structures that decay into a Hartmann flow configuration further downstream.
- *Quasi-2D flow with Shercliff wall jets* (QM). Q2D structures are initially formed, but are replaced by buoyancy-driven streamwise side jets once the Q2D structures decay. The jets form an M-shaped profile.
- *Quasi-2D flow with buoyancy-driven backflow jets* (QW). The duct is populated by Q2D structures which are replaced by buoyancy-driven backflow side jets once the Q2D structures decay.
- *Unstable flow with sidwalls jets* (UL). The distribution of electrical currents forces the flow to the Shercliff layers. At low $Ha$ these jets are unstable due to the presence of vortex promoters, and at high $Ha$ they are linearly unstable.

We describe each of these flows in detail in sections 3.3 to 3.6.

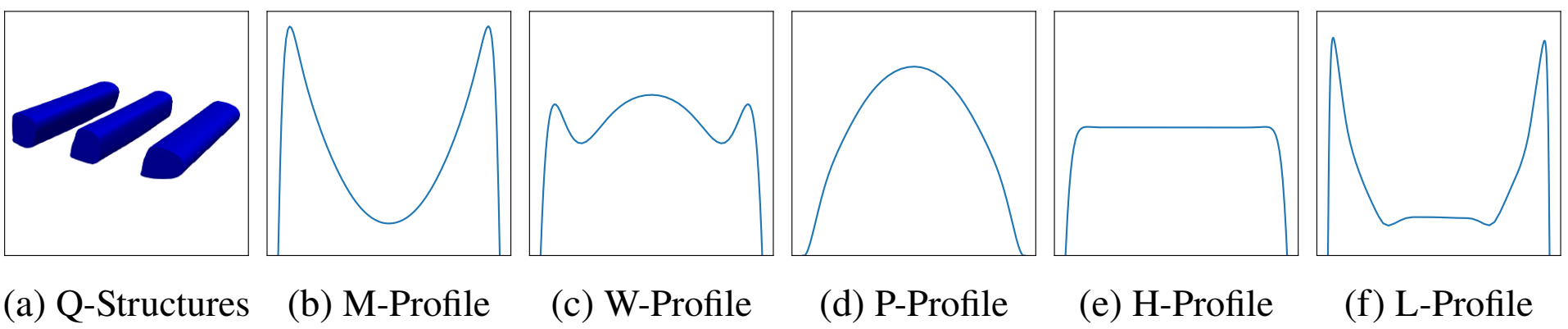

(a) Q-Structures (b) M-Profile (c) W-Profile (d) P-Profile (e) H-Profile (f) L-Profile

Figure 4. Representative shape of the structures represented by each letter from table 2, with vertical velocity isosurfaces (a) and time-averaged velocity profiles between Shercliff walls in panels (b)-(f).

For the values of $Ha$ and $Re$ we have considered, the flow is determined by the buoyancy force ($F_b$) and the Lorenz force ($F_L$). For the buoyancy force specifically we consider either horizontal ducts or vertical ducts, so it acts only in either the streamwise $x$- or vertical $z$-direction (see figure 2). With the duct in the vertical position, there is a significant change to the flow structure. In the horizontal case, the changes are much less significant so we will only mention them in passing.

For each of the 4 flow types, we compare their average velocities, Reynolds stresses (defined shortly), turbulent kinetic energy and Nusselt number. For cases where $Gr = 0$, we still solve equation 2.3, which results in energy being transported as a passive scalar. This allows us to quantify the effect of the buoyancy force, and we did not observe meaningful differences in velocity profiles, statistical moments, TKE or $Nu$ when heat transfer is added to ducts with a high wall conductance ratio.

We apply the Reynolds decomposition to some quantity $\chi$ as $\chi = \overline{\chi} + \chi'$ where $\overline{\chi}$ is the temporal mean and $\chi'$ the fluctuations and compute the Reynolds stresses. Henceforth, we will let $u$ denote the streamwise, $v$ the spanwise and $w$ the vertical components of the velocity. In all cases, the spanwise component is several orders of magnitude smaller than the other two due to the effects of the magnetic field, so we will consider it only when computing the TKE, which is denoted as $E_{\text{kin}}$ and given by

$$E_{\text{kin}}(\boldsymbol{x}) = \frac{1}{2}\left(\overline{u'u'} + \overline{v'v'} + \overline{w'w'}\right) \tag{3.1}$$

We will also compute the time evolution of the volume-averaged $E_{\text{kin}}$, which is given by

$$\langle E_{\text{kin}}(t)\rangle_V = \frac{1}{V}\int_V E_{\text{kin}}(\boldsymbol{x}, t)dxdydz. \tag{3.2}$$

The DNS conducted for this part were performed for 600 convective time units, corresponding to 6 full turnover times for the flow in the duct. The averaging was done over the last 400 convective time units which is enough to ensure that the flow is already fully developed.

In the second part of our exploration, we perform a parametric study to find the transition between the different regimes on the $F_b^x - c_W$ plane at a fixed $Ha = 325$ and $Re = 4000$. These specific values have the shortest transients before the flow is fully developed, and therefore we can classify the flow from the different simulations after a shorter simulation, specifically we use only 150 convective time units. For some combinations of $F_b^x$ and $c_W$ where the flow is at a transition between two different types, we extended our simulations for another 150 convective time units. This is duly noted when applicable.

### 3.2. *The naming convention*

Due to the large amount of distinct phenomena we have observed, it is necessary to define a concise means of classification that clearly distinguishes each flow. To this end, we identify each type of flow by two letters. Each one represents a characteristic phenomena observed in the flow, and can be thought as the ”building blocks” of the overall flow. Table 2 lists the 7 letters we use, describes the type of flow, and figure 4 provides an image of a typical example of the phenomena the letter represents. Since the duct considered in our simulations is very long, most of the flows we have observed involve a transition from one state to another. For example, a QH flow is a flow that transitions from Q2D rolls, as discussed in Krasnov *et al.* (2023*b*) to a Hartmann flow towards the end of the duct. In some cases, both regimes are unstable and their interaction yields interesting results, this is the case of for example the QM flow discussed in section 3.5. Finally, in some cases with highly conducting duct walls, the flow is the same along the entire duct. In those cases we specify if it is stable or unstable, such as the UL flow discussed in section 3.4.

| Letter | Description of phenomena |
|---|---|
| $Q$ | Quasi-2D Rolls in the flow |
| $M$ | Buoyancy-driven jets at the Shercliff walls for upward flow. These are typically weaker and wider than jets driven by the Lorenz force in the Hunt or Walker flows (see $L$ type below) |
| $W$ | Buoyancy-driven dips at the Shercliff layer for downward flow. In extreme cases, the dips can cause counterflow jets |
| $P$ | Poiseuille-like flow |
| $H$ | Hartmann flow |
| $L$ | Side jets by the Lorenz force (Hunt or Walker flow) |
| $S$ | Linearly stable flow |
| $U$ | Linearly unstable flow |

Table 2. Naming convention employed to classify flow types.

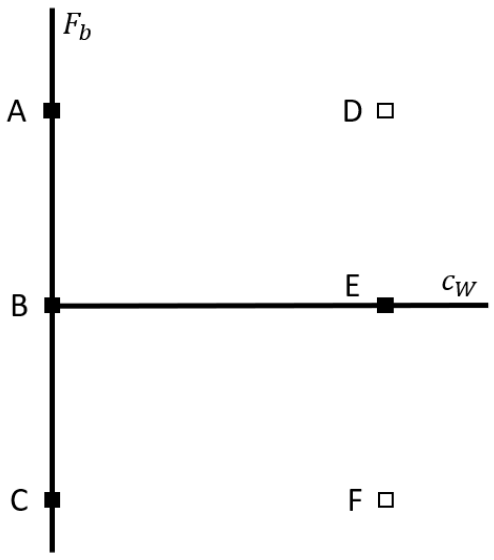

Figure 5. Simulation parameters projected on the $c_W - F_b$ plane. Each simulation was conducted at four levels of $Ha$, 325, 500, 750 and 1000. Solid squares represent runs whose results we show in this paper, empty squares represent simulations the results of which are virtually identical to the $F_b = 0$ and not shown.

### 3.3. *Horizontal duct with insulating walls: The QH flow*

For the simulations conducted in horizontal electrically insulating ducts, the inlet jets promote the generation of vortices which were elongated in the magnetic field direction and advected along the duct. Eventually, these vortices decay into a Hartmann flow. Following the naming convention established in section 3.2, we call this flow the QH flow. This occurs for all combinations of $Re$ and $Ha$ we explored, both with and without heat transfer. When heat transfer is present, the 2D rolls appear even without the vortex promoters, but instead of forming at the inlet they appear further downstream. In shorter ducts, these buoyancy-driven Q2D vortices do not form and the flow remains Hartmann-like.

| $Re$ | $Ha$ | $Gr$ | $Pr$ | Duct Orientation | Streamwise Direction | $c_W$ |
|---|---|---|---|---|---|---|
| 4000 | 325 | 0 | 0.02 | NA | NA | 0 |
| 4000 | 500 | 0 | 0.02 | NA | NA | 0 |
| 4000 | 750 | 0 | 0.02 | NA | NA | 0 |
| 4000 | 1000 | 0 | 0.02 | NA | NA | 0 |
| 4000 | 325 | 0 | 0.02 | NA | NA | 0.1 |
| 4000 | 500 | 0 | 0.02 | NA | NA | 0.1 |
| 4000 | 750 | 0 | 0.02 | NA | NA | 0.1 |
| 4000 | 1000 | 0 | 0.02 | NA | NA | 0.1 |
| 4000 | 325 | $5 \times 10^7$ | 0.02 | Vertical | Upwards | 0 |
| 4000 | 500 | $5 \times 10^7$ | 0.02 | Vertical | Upwards | 0 |
| 4000 | 750 | $5 \times 10^7$ | 0.02 | Vertical | Upwards | 0 |
| 4000 | 1000 | $5 \times 10^7$ | 0.02 | Vertical | Upwards | 0 |
| 4000 | 325 | $5 \times 10^7$ | 0.02 | Vertical | Downwards | 0 |
| 4000 | 500 | $5 \times 10^7$ | 0.02 | Vertical | Downwards | 0 |
| 4000 | 750 | $5 \times 10^7$ | 0.02 | Vertical | Downwards | 0 |
| 4000 | 1000 | $5 \times 10^7$ | 0.02 | Vertical | Downwards | 0 |

Table 3. Parameters for simulations in sections 3 and 4. For distinct $Ha$ are used. For the insulating wall case the only effect is to dissipate disturbances, however with conducting walls higher $Ha$ induces instability. $Gr$ is either 0 or $10^7$, and $c_W$ wither 0 or 0.1. In section 6, intermediate values of $Gr$ and $c_W$ will also be considered.

The decay of the vortices occurs faster as $Ha$ is increased, due to the larger Lorenz force. This is clearly visible in the patterns of the streamwise and vertical velocities in the $(x, z)$-midplane (figure 6) and by all three centerline velocity components (figure 7), shown for the two extreme values of $Ha$ at $Re = 4000$.

The resulting Reynolds stresses from these Q2D rolls resemble those of 2D turbulence (Jiménez 1990). Figures 8 and 9 show the time averaged streamwise velocity profile between the Shercliff walls and the relevant components of the Reynolds stress tensor between those same walls. The transition form Q2D rolls to Hartmann flow is especially visible at $Ha = 1000$, whereas at $Ha = 325$ the decay is not yet complete. The time averaged vertical component of velocity is close to zero, and the remaining components of the Reynolds stress tensor are several orders of magnitude smaller than those associated with the streamwise and vertical velocity components, so we only include those in the figures.

As $Ha$ increases, it takes longer for the flow to fully develop. At the upper limit of our simulations, $Ha = 1000$ the flow had two intermediate states. First, each of the jets developed its own set of quasi-2D vortices. Eventually those decay and the flow becomes completely laminar for a short while. Then temporal instabilities close to the inlet trigger the appearance of only one streak of vortices.

The lifespan of the double vortex streak can be shortened or even suppressed entirely if instabilities are added into the system, for example in the form of random noise at the inlet or the addition of heat transfer between the Shercliff walls. Figures 10 and 11 show the effects of adding heating in a horizontal duct. The effect is most visible downstream, when the initial rolls have dissipated and are replaced by buoyancy-driven vortices. These occur naturally even without vortex promoters, but are attached to the lower wall which only improves heat transfer at one side.

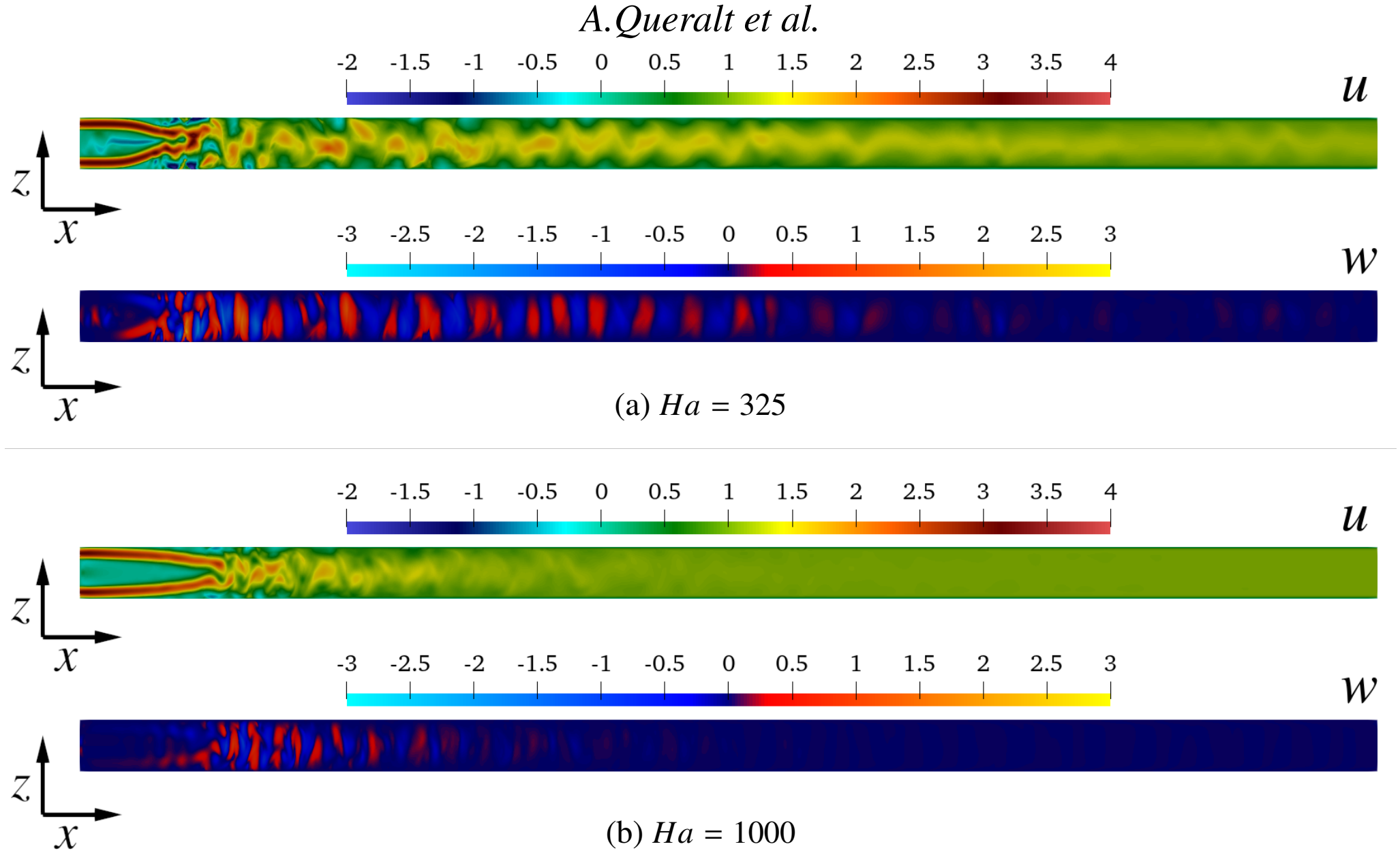

(a) $Ha$ = 325

(b) $Ha$ = 1000

Figure 6. Instantaneous snapshots of the streamwise $u$ and vertical $w$ velocities at the duct $(x, z)$-midplane $y = 0$, for $Re = 4000$, $Gr = 0$ in perfectly insulating ducts $c_W$=0. Ducts have been rescaled by a factor of 0.5 in the $x$ direction to fit the page.

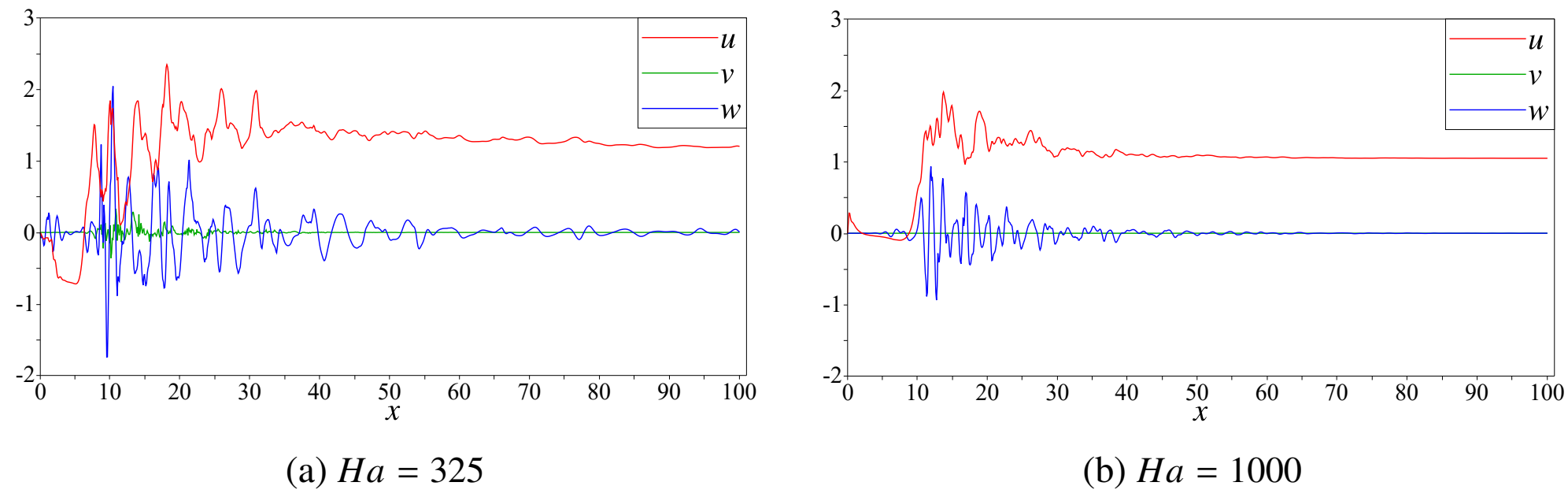

(a) $Ha$ = 325 (b) $Ha$ = 1000

Figure 7. Velocity at the duct centerline at $Re = 4000$, $c_W = 0$, $Gr = 0$ and different values of $Ha$. The decay of the Q2D structures is observed to happen sooner at higher $Ha$.

### 3.4. *Duct with highly conducting walls: The UL flow*

When the walls are highly conducting (points C, D and E in figure 5), an unstable Walker flow with jet detachments (UL as per our naming convention) populates the duct. We differentiate two sources of instability, (*i*) those induced by the vortex promoters and (*ii*) those induced naturally by the inflection points of the jets close to the Shercliff walls. The difference between the two is clearly visible in figure 12. In the former case, instabilities are triggered only up to $Ha = 500$. From $Ha = 750$ on, the Lorenz force is strong enough to completely dampen the inlet instabilities. In these cases the flow is unstable even without vortex promoters. Figure 13 shows the centreline velocity at four $Ha$ for $Re = 4000$. Initially the duct is completely populated by promoter driven instabilities, but those become less frequent as $Ha$ is increased. When the naturally occurring instabilities appear their magnitude is smaller than that of the ones created by the vortex promoters. The naturally driven instabilities appear after a critical duct length $L_c$, which we have observed to get smaller as $Ha$ increases. We attribute this to the side jets getting thinner as $Ha$ increases, which steepens the velocity gradients near the Sheriff walls and creates stronger inflection points in the velocity profile. While we have failed to see any qualitative difference in the flow when adding heat transfer, there is some quantitative difference in the $\langle E_{\mathrm{kin}} \rangle_V$. Is essentially random noise with highly conducting walls. Because the Lorenz force is proportional to the magnitude of the velocity, the buoyancy effects are compensated by a change in the Lorenz force. At higher $Gr$ or with a weaker Lorenz force a qualitative difference can be observed. We discuss this further in section 6. Within the parameter range we have considered, the mean, bulk, and wall temperatures of the fluid behave as if they were passively transported in the flow, regardless of duct orientation.

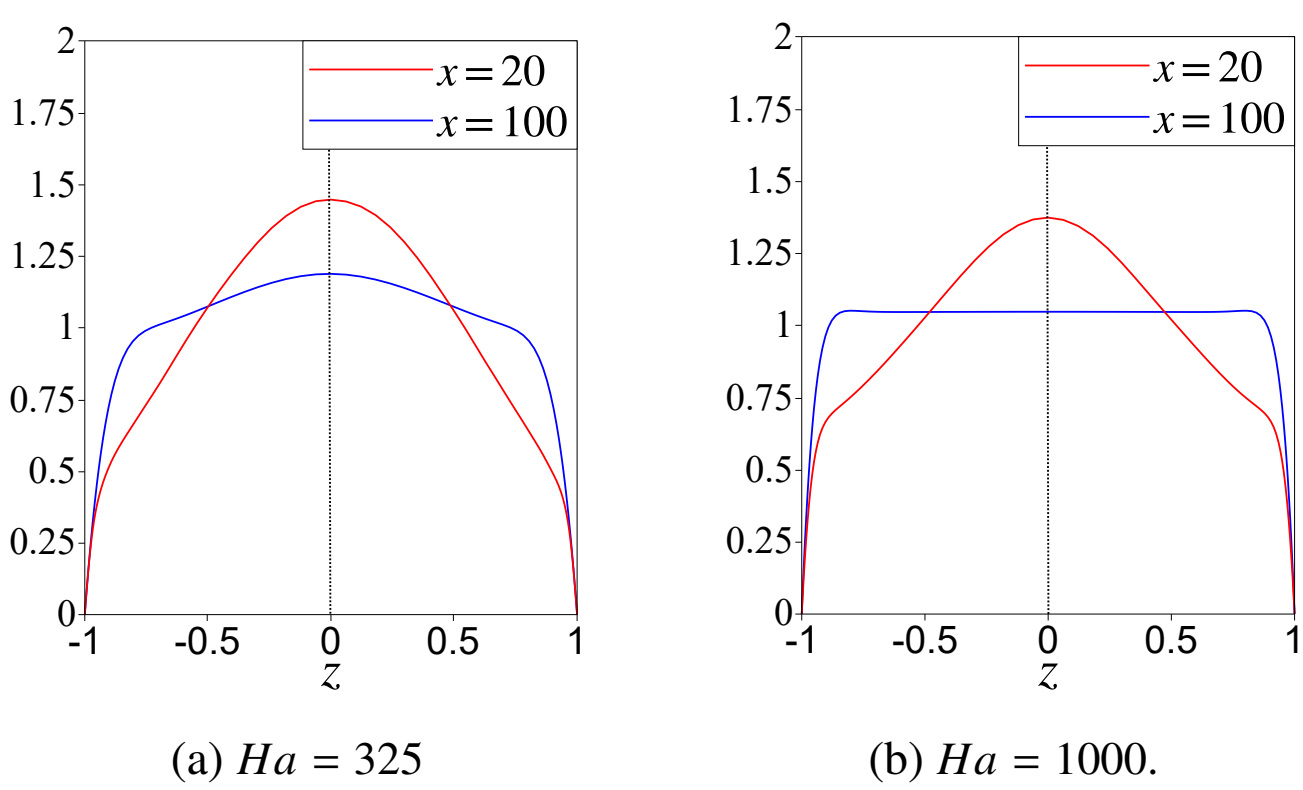

(a) $Ha$ = 325 (b) $Ha$ = 1000.

Figure 8. Time-averaged streamwise velocity profiles between Shercliff walls at two duct cross sections at $Re$ = 4000, $c_W$ = 0 and different values of $Ha$. The duct segments with Q2D structures present a 2D turbulence profile, once they decay the profile is that of a Hartmann flow.

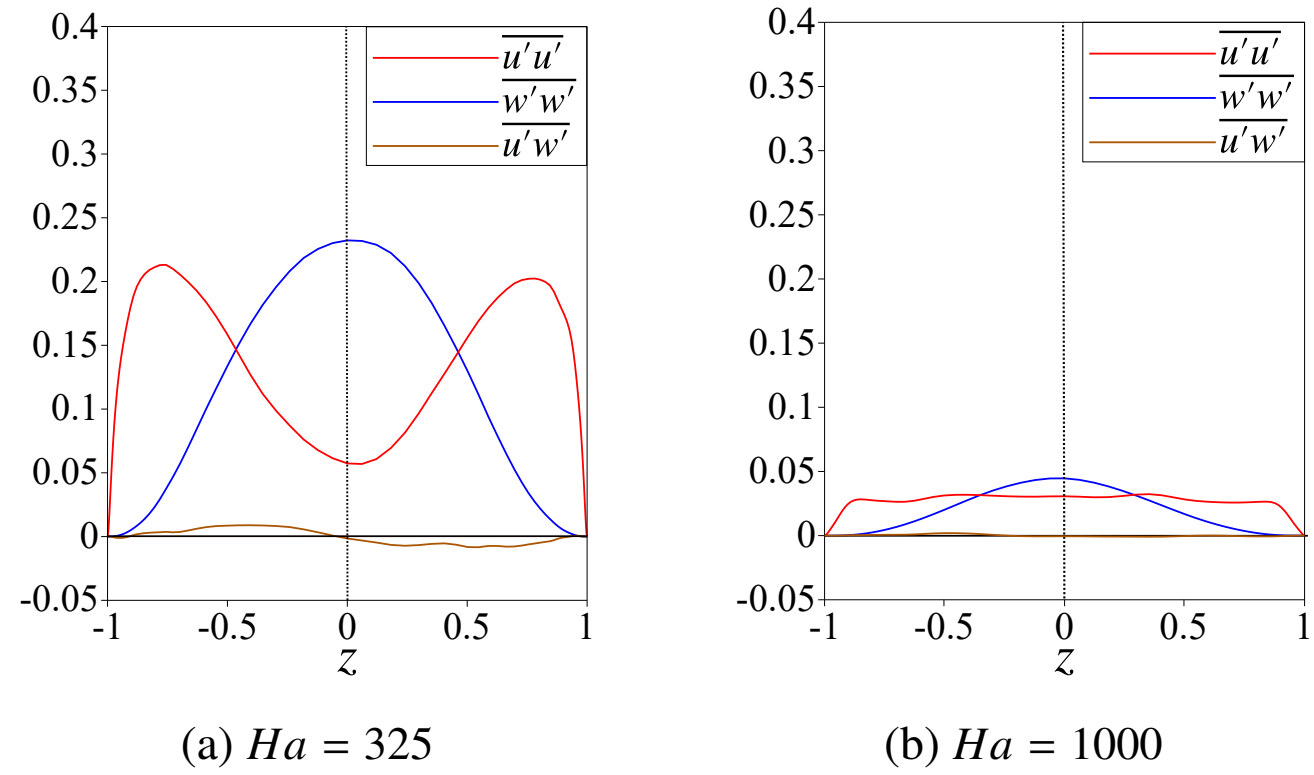

(a) $Ha$ = 325 (b) $Ha$ = 1000

Figure 9. Reynolds stresses between Shercliff walls at $x$ = 30. The components are given in the legend.

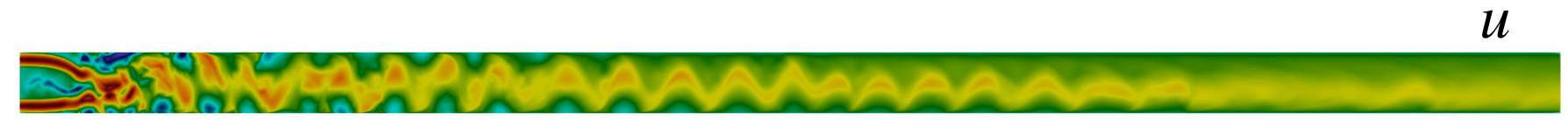

Figure 10. Streamwise velocity $u$ at $(x, z)$-midplane at $Re$ = 4000, $Ha$ = 325 and $Gr = 10^7$ with the duct in a horizontal position, applying symmetric heating.

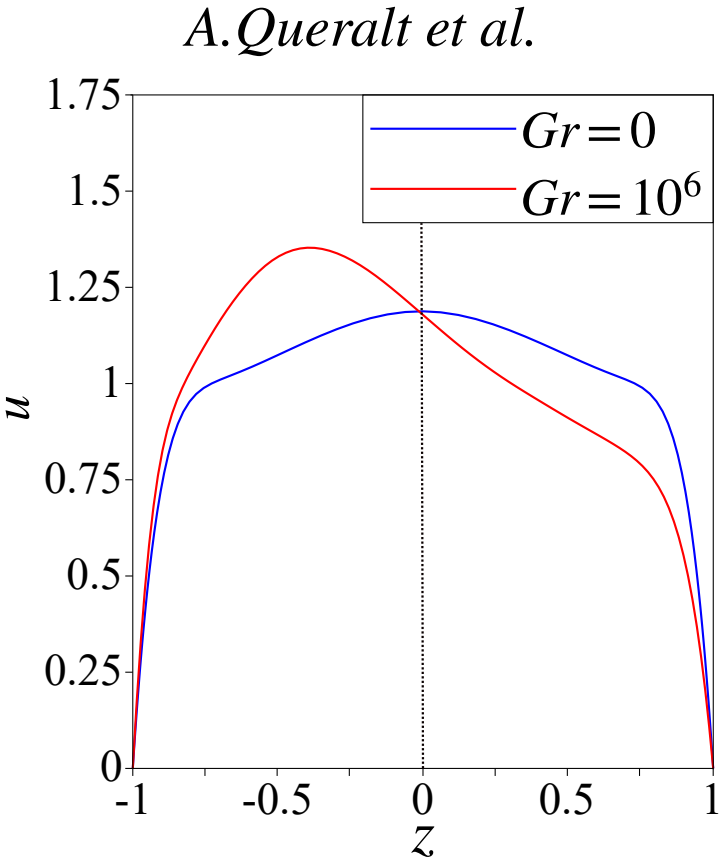

Figure 11. Time-averaged streamwise velocity profiles between Shercliff layers at $x = 100$, with (red) and without (blue) heat transfer in a horizontal duct. The heat flux casues 2D vortices to naturally appear attached to the lower wall of the duct. These also appear here and cause an assimetry in the velocity field.

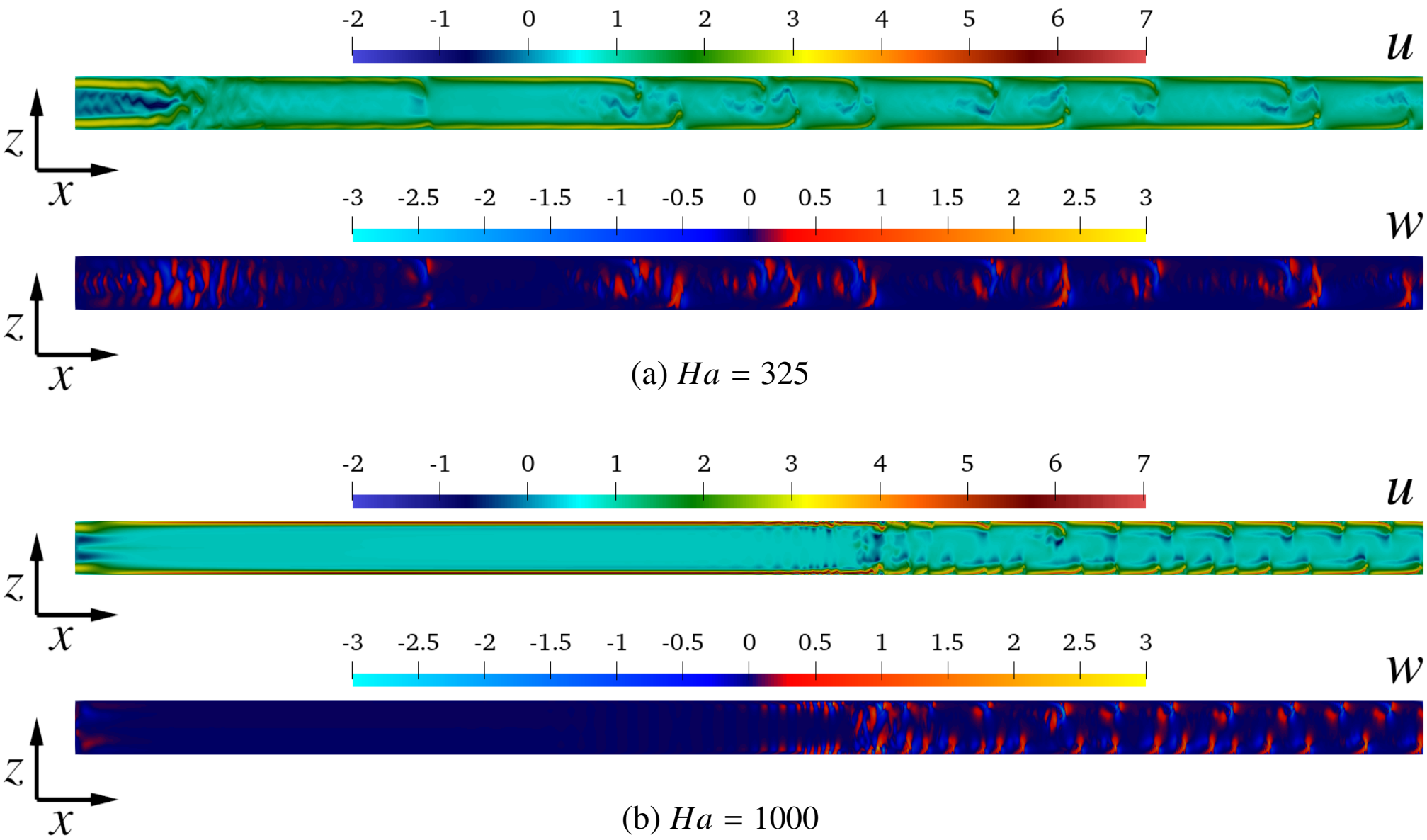

Figure 12. Instantaneous snapshot of the streamwise $u$ and vertical $w$ velocity components at $Re = 4000$, $Gr = 0$, $c_W = 0.1$, ducts have been rescaled by a factor of 0.5 in the $x$ direction to fit the page.

The strong vorticity at the side walls, specially at higher $Ha$ causes small vortices to appear at the Shercliff layer. Counter rotating vortices also appear in the bulk of the flow attached to the Shercliff layer. Some of these vortices grow and advect the metal from the Shercliff layer to the core, creating the detachments. Figure 14 shows this process Here only the duct segment from $x = 45$ to $x = 85$ is shown. A Line Integral Convolution (LIC, see, e.g.Cabral & Leedom (1993)) is used to visualize the velocity fluctuations field. The initial vorticities that precede the detachments are visible at the left side of the figure. The first detachments occur in the middle of the figure, where the vortices from both side walls interact. The Lorenz force dampens the fluctuations at the centre of the duct, so detachments from both walls don't merge into single vortices, as will happen in sections 3.5 and 3.6.

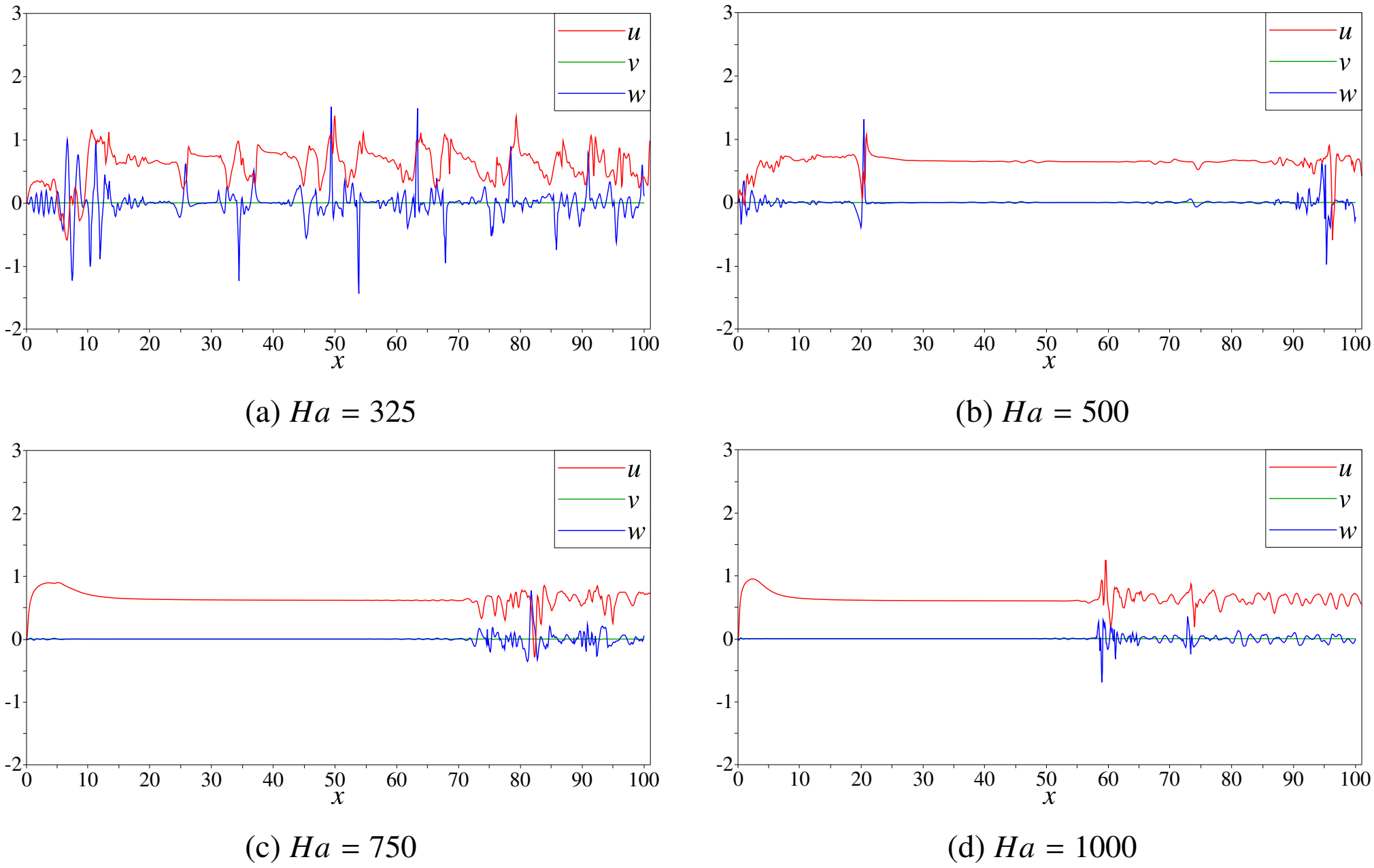

(a) $Ha = 325$

(b) $Ha = 500$

(c) $Ha = 750$

(d) $Ha = 1000$

Figure 13. Velocity along the duct centerline at $Re = 4000$, $c_W = 0.1$, $Gr = 0$ and different values of $Ha$ in (a) the instabilities are driven by the vortex promoters, in (c) and (d) they arise naturally and in (b) both driven and natural instabilities are present.

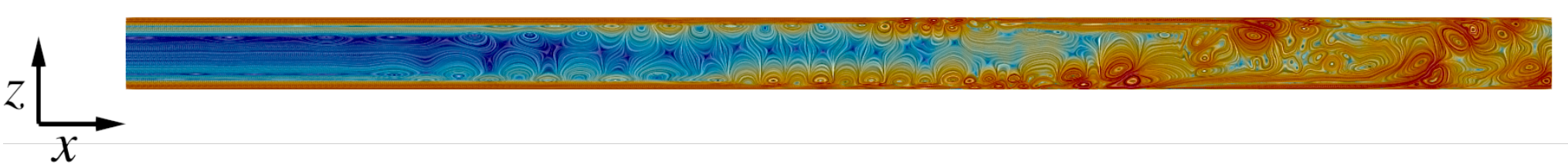

Figure 14. Velocity fluctuations magnitude at the $(x, z)$ midplane segment from $x = 45$ to $x = 85$, $Ha = 750$, $c_W = 0.1$. Subdomain is shown in original dimensions without any rescaling. Two vortex streaks at the Shercliff layers grow until they become unstable.

The time-averaged velocity profiles between Shercliff layers (figure 15) show that in the region populated by jet detachments the jets themselves become thicker. As a consequence, the wall-normal velocity gradients will be much smaller. Figure 16 shows the Reynolds stresses between the Scherliff walls also at different duct locations. For the naturally occurring detachments, the highest stresses are obtained just before the detachments occur. In all cases, whenever there are detachments, the streamwise component of the stresses has a pronounced $M$ shape. For the $z$ component $\overline{w'w'}$, however, the shape is different for natural and promoted detachments, the former having an $M$ shape and the latter having a maxima at the centre and decaying at either side.

The presence of two mechanisms that promote instability in the highly conducting wall case. By artificially adding more vorticity to the flow, (i.e making the side jets thinner), which in a practical context would be equivalent to placing a larger cylinder as an obstacle at the duct inlet, one could potentially increase the value of $Ha$ at which instabilities are triggered from the onset of the flow. In addition, this could increase the number of instabilities for parameter combinations that have very few of them. A complete study at intermediate $Ha$, with different inlet jet sizes and without any jets at all is needed to study these phenomena in depth.

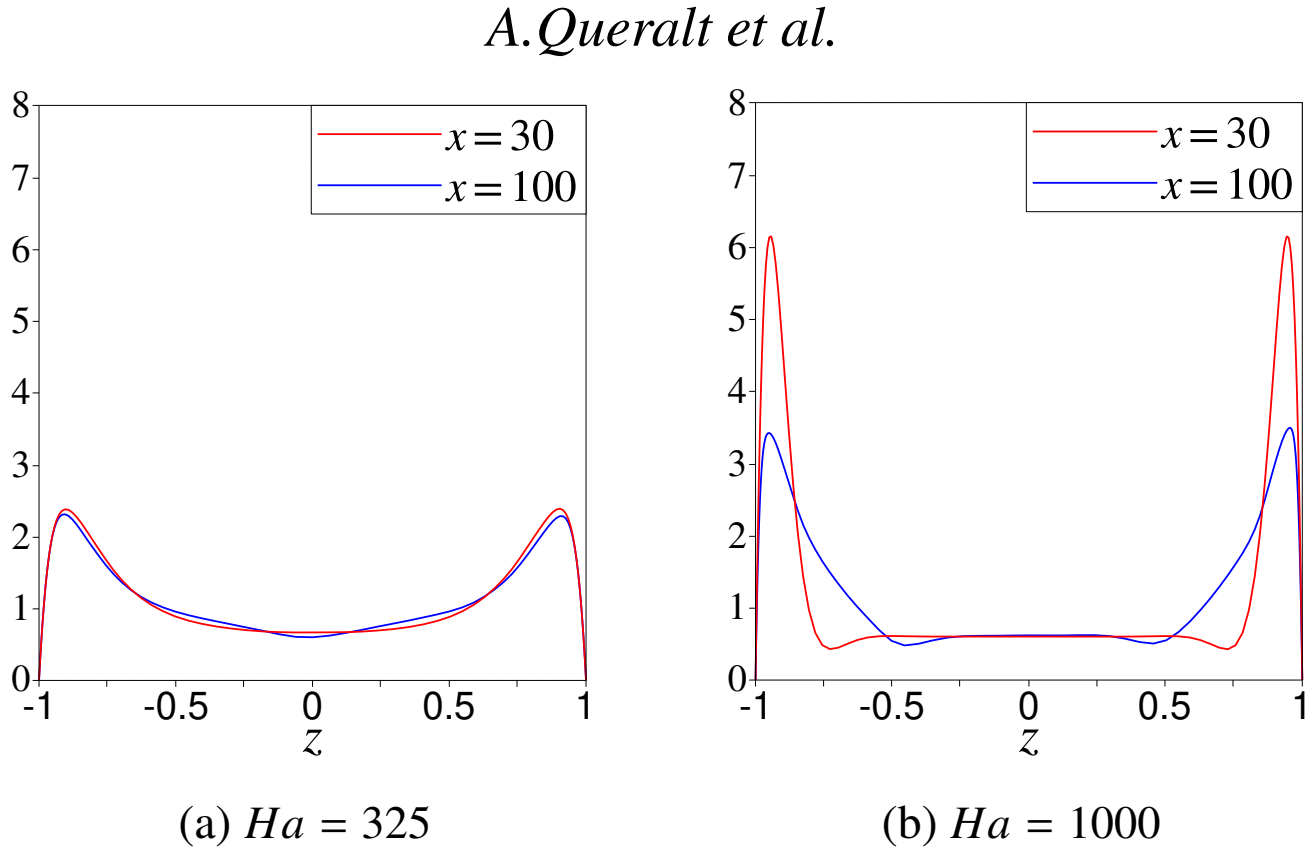

(a) $Ha$ = 325 (b) $Ha$ = 1000

Figure 15. Time-averaged streamwise velocities between Shercliff walls at different duct cross sections. At $Re$ = 4000, $Gr$ = 0, $c_W$ = 0.1 and different values of $Ha$. Two distinct profiles arise when there are naturaly promoted instabilities in the duct, whereas the mean velocity is constant in the duct where instabilities are triggered by vortex promoters.

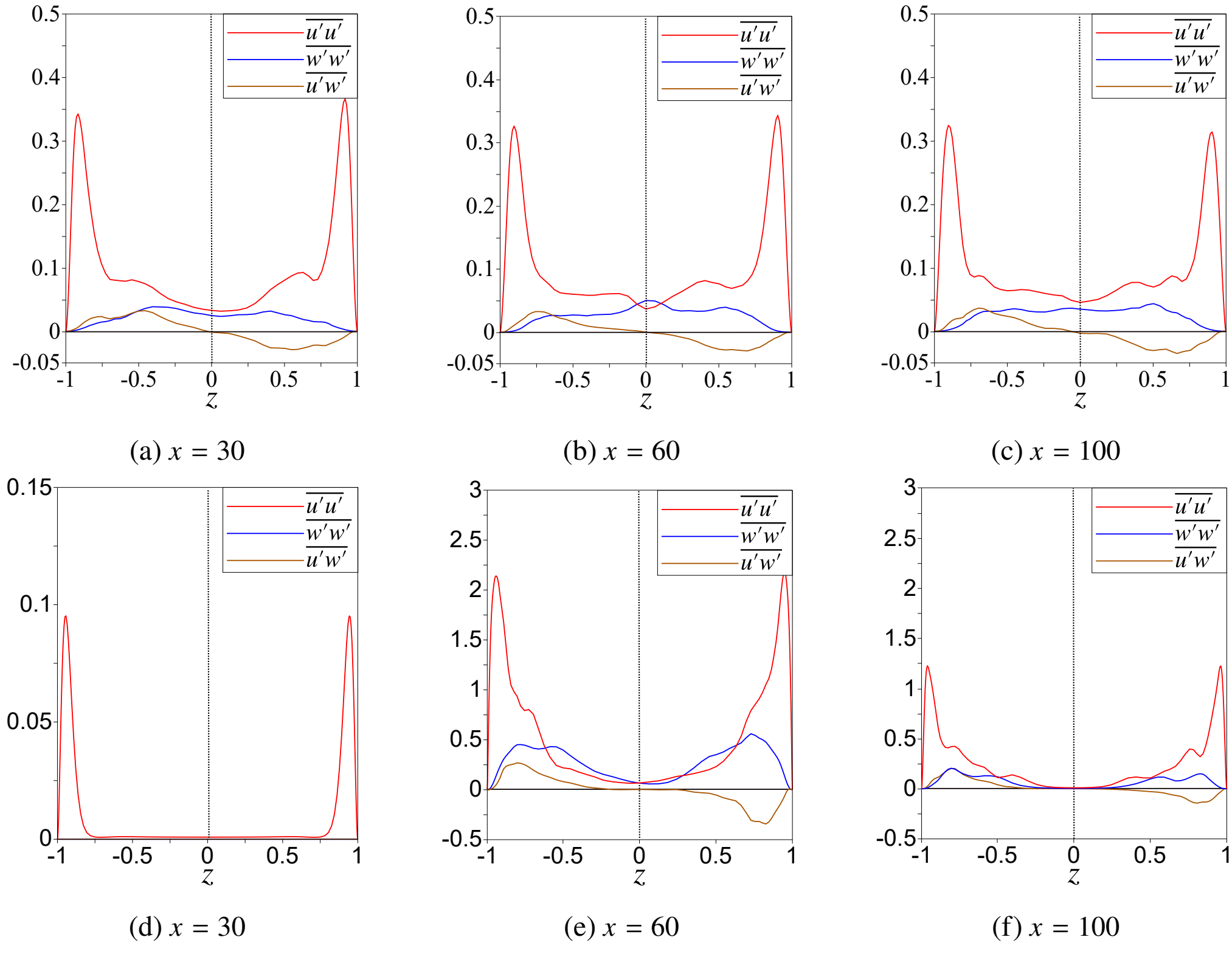

(a) $x$ = 30 (b) $x$ = 60 (c) $x$ = 100

(d) $x$ = 30 (e) $x$ = 60 (f) $x$ = 100

Figure 16. Reynolds stress components $\overline{u'u'}$, $\overline{w'w'}$ and $\overline{u'w'}$ at different $x$ coordinates for $Re$ = 4000, $Gr$ = 0, $c_W$ = 0.1 and $Ha$ = 325 $(a - c)$, $Ha$ = 1000 $(d - f)$. The spanwise component $\overline{v'v'}$ is several orders of magnitude smaller and not pictured. Note the scale is different in sub figure (a) for readability. The stresses present a peak at the point where the detachments first occur $x \approx 60$.

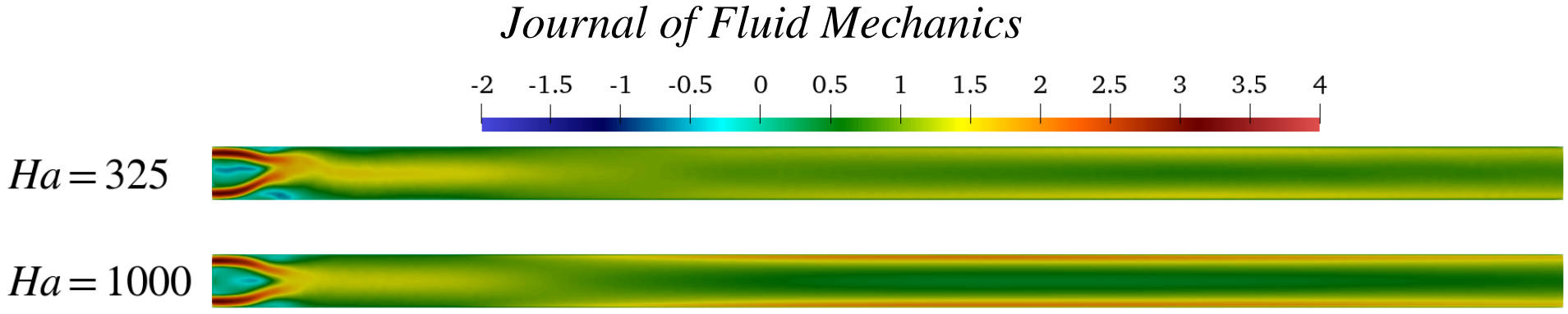

Figure 17. Time-averaged streamwise velocity component for an upwards flow in a perfectly insulating duct $c_W = 0$ at $Ha = 325$ and 1000 and $Re = 4000$.

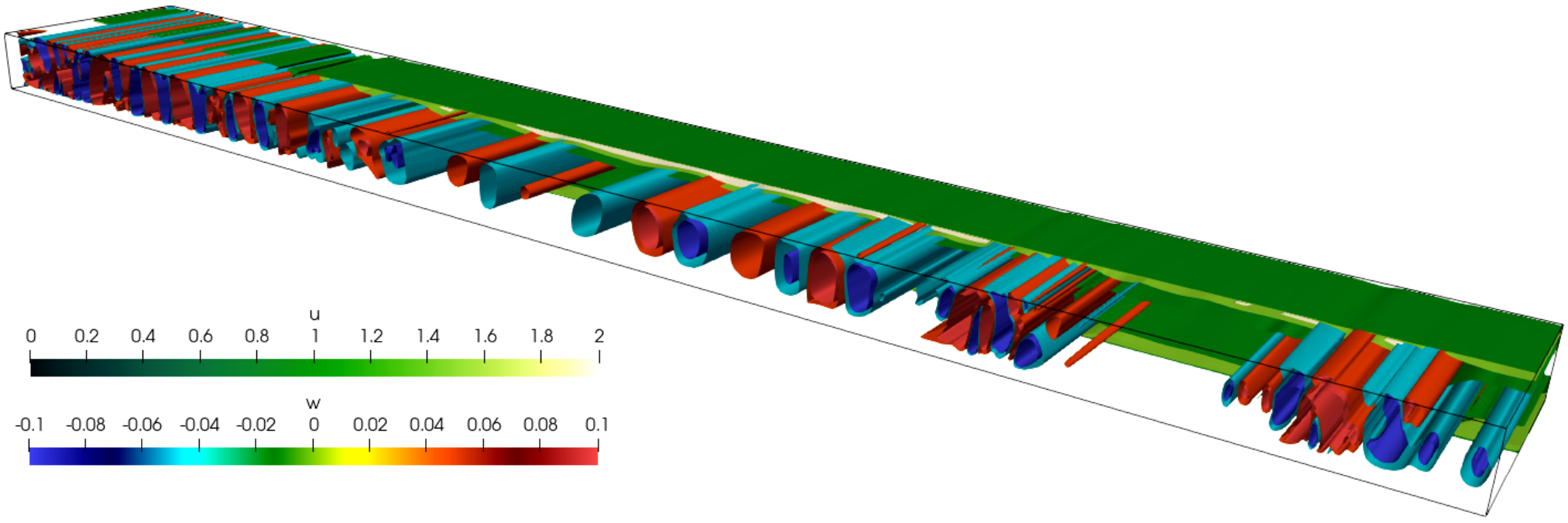

Figure 18. Isosurfaces for the streamwise velocity $u$ (green) and vertical velocity $w$ (red-blue) for $Re = 4000$, $Ha = 1000$, $Gr = 10^7$ and $c_W = 0$ with the duct in a vertical position and upwards flow.

### 3.5. *Perfectly insulating, vertical duct with upwards flow: The QM case*

In perfectly insulating ducts, when the buoyancy force acts in the direction of the mean flow (point $A$ in figure 5), the flow initially retains the Q2D rolls present in the horizontal case. Once the rolls decay, the flow becomes dominated by buoyancy driven side jets. As with the UL flow, the increased vorticity creates vortices attached to the Shercliff walls, however, unlike the UL flow, when the jets detach, the two vortices merge together re-establishing rolls whose characteristic scale is that of the duct. Figure 17 shows the time-averaged streamwise velocity component where the transition between both flow regimes is clearly noticeable. The Q2D rolls, (both those created by inlet disturbances and by the side jets) exist "sandwiched" between the two side jets. Figure 18 shows vertical and streamwise velocity isosurfaces, where the presence of the initial Q2D rolls, the jet-driven Q2D rolls and gaps between roll clusters is visible. These gaps are regions where the original rolls have already dissipated but have not been replaced by jet-driven rolls. The roll-triggering process occurs intermittently, and only when the vortex promoters are present.

This secondary roll-generation is a source of intermittency for the overall flow. This means that the number of vortices inside the duct changes in time. This is specially true at higher $Ha$, where the original vortices are dissipated much sooner. At the lower end of our $Ha$ range however, despite intermittency also being present, the Q2D rolls are present across the entire duct, although they are not as uniform as those present in the QH flow. This is evident from figure 19 which show the midplane vertical and streamwise velocities, and from figure 20, which shows the centerline velocities at different $Ha$. Compared to the same profile for the QH (figure 8) it becomes apparent that positioning the duct in an

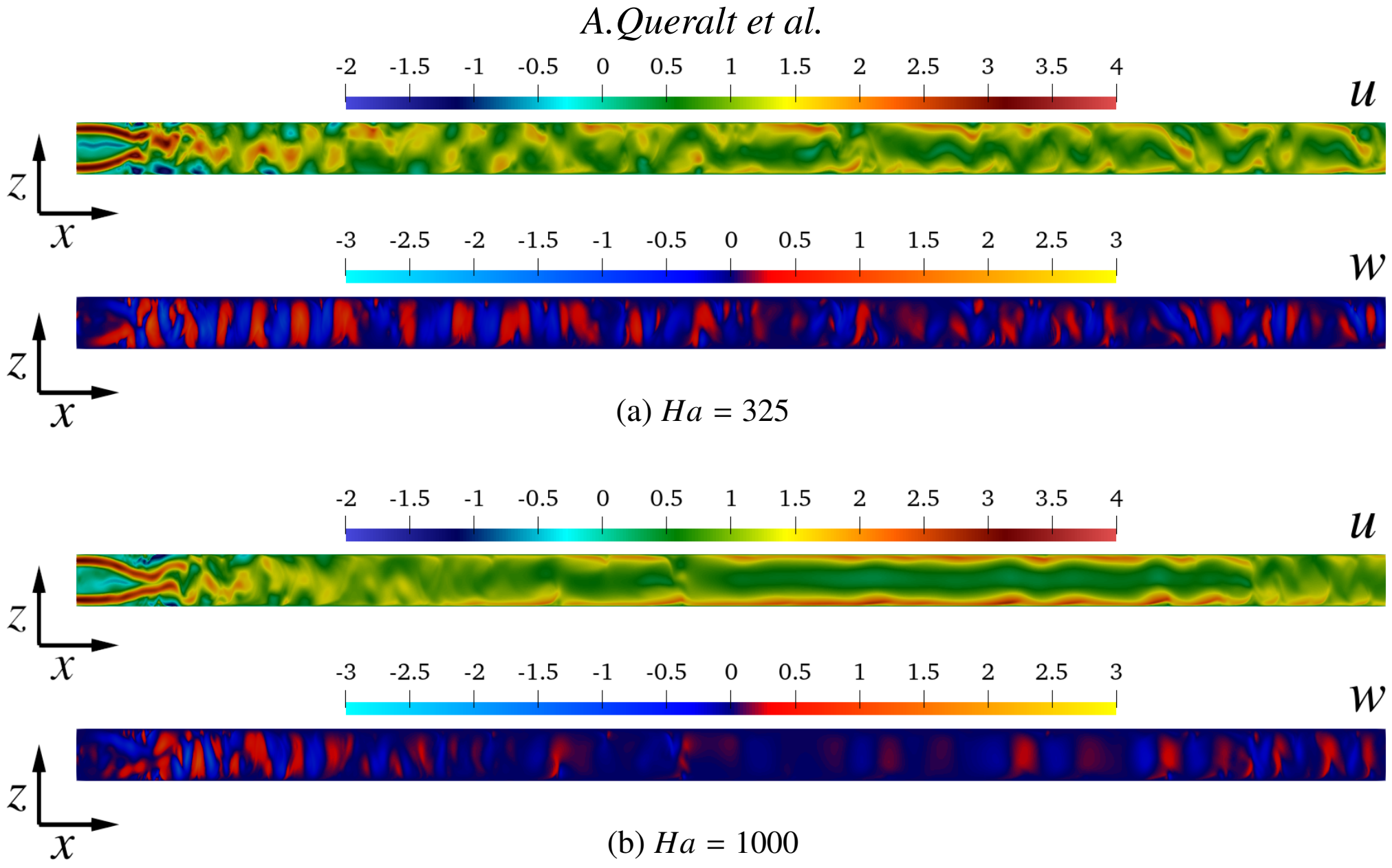

(a) $Ha$ = 325

(b) $Ha$ = 1000

Figure 19. Instantaneous snapshot of the streamwise $u$ and vertical $w$ velocity components at $Re$ = 4000, $Gr$ = $10^7$, $c_W$ = 0 in a vertical duct with upwards flow. Ducts have been rescaled by a factor of 0.5 in the $x$ direction to fit the page.

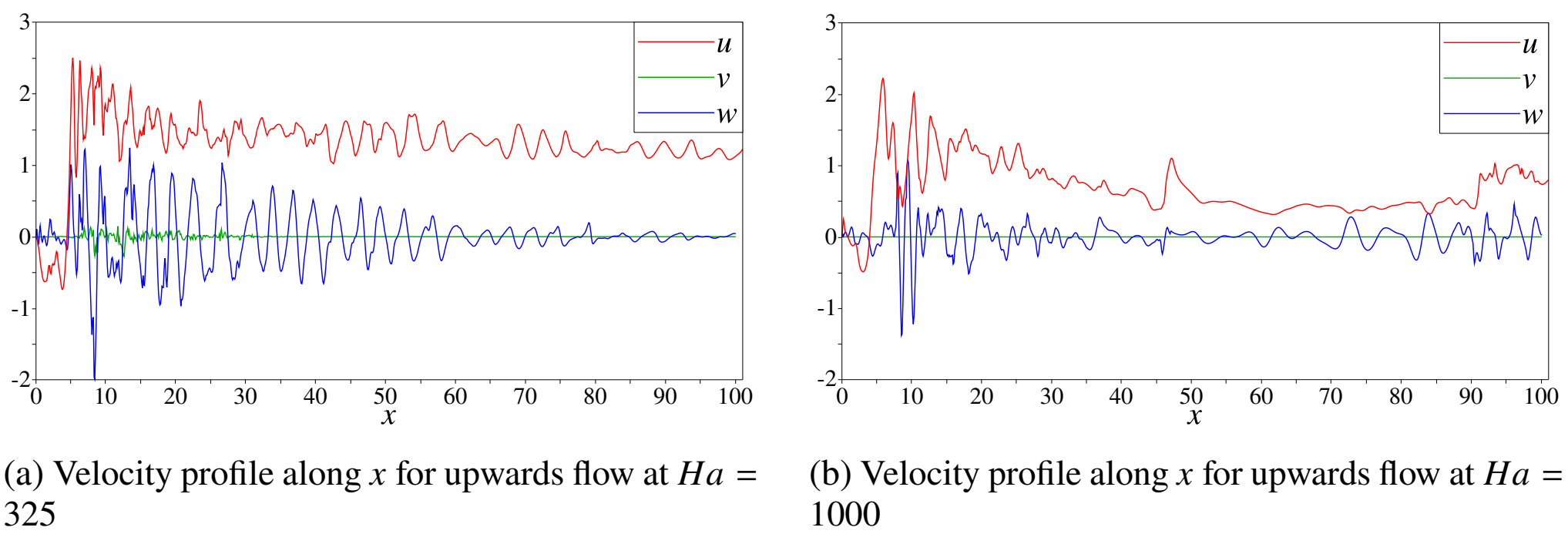

(a) Velocity profile along $x$ for upwards flow at $Ha$ = 325

(b) Velocity profile along $x$ for upwards flow at $Ha$ = 1000

Figure 20. Velocities at the duct centre, upwards flow at $Re$ = 4000, $Gr$ = $10^7$, $c_W$ = 0 and different $Ha$.

upwards orientation succeeds in extending the life of the vortices, specially at the lower end of the $Ha$ range.

The two-dimensional turbulent stresses (Jiménez 1990) we observed in the QH flow are also present in the QM flow, but only for the region before the side jets appear. figures 21 and 22 show the time-averaged streamwise velocity and turbulent stresses between the Shercliff walls at three positions along the duct: The Q2D region at the beginning ($i$), the transition region that on average takes a Hartmann line profile ($ii$) this transitions occurs at different $x$ coordinates depending on $Ha$. And the buoyancy jet dominated region ($iii$). From a qualitative point of view most can be said to have a 2D turbulence stress profile, however, at higher $Ha$ the $M$ shape of the streamwise stress component is sharper, and in the buoyancy jet region the profile presents a third peak at the centre of the duct at low $Ha$.

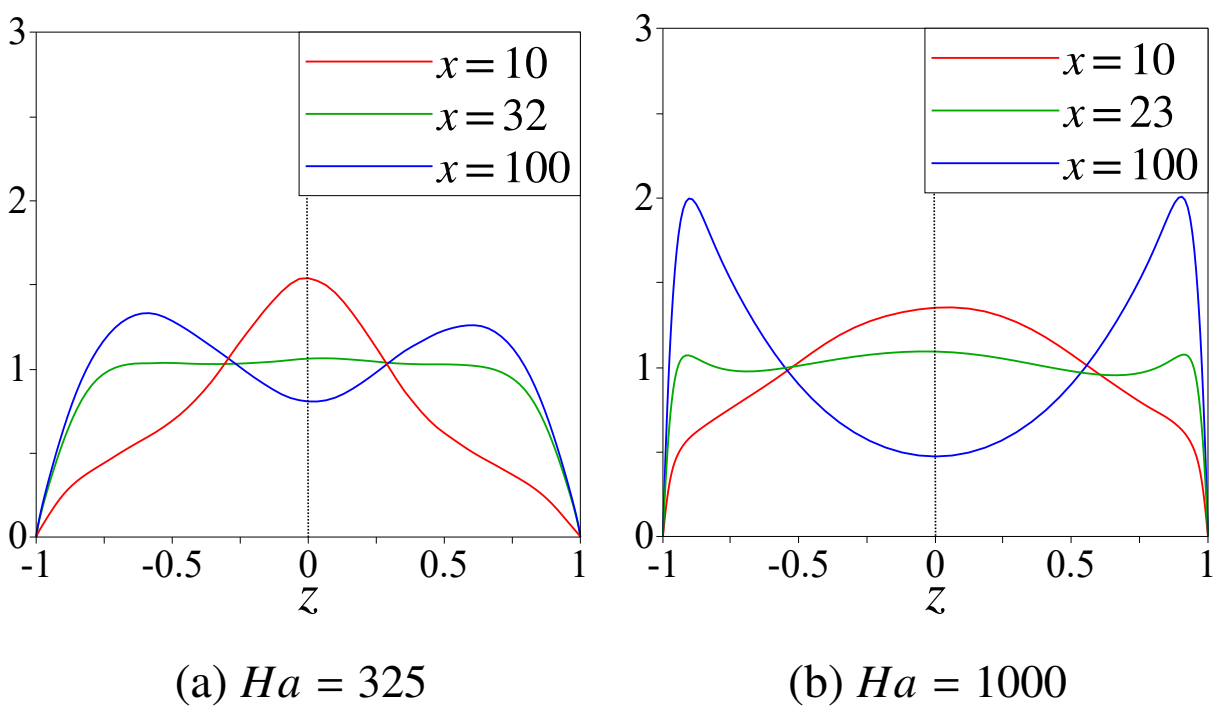

(a) $Ha$ = 325 (b) $Ha$ = 1000

Figure 21. Time-averaged streamwise velocities between Shercliff walls for the upwards QM-flow at $Re$ = 4000, $c_W$ = 0, $Gr$ = $10^7$ and $Ha$ = 325 and 1000. Profiles at $x$ coordinates with Q2D structures (red at $x$ = 10), transition to Hartmann-like profile (green, different for each $Ha$) and fully developed buoyancy driven side jets (blue at $x$ = 100).

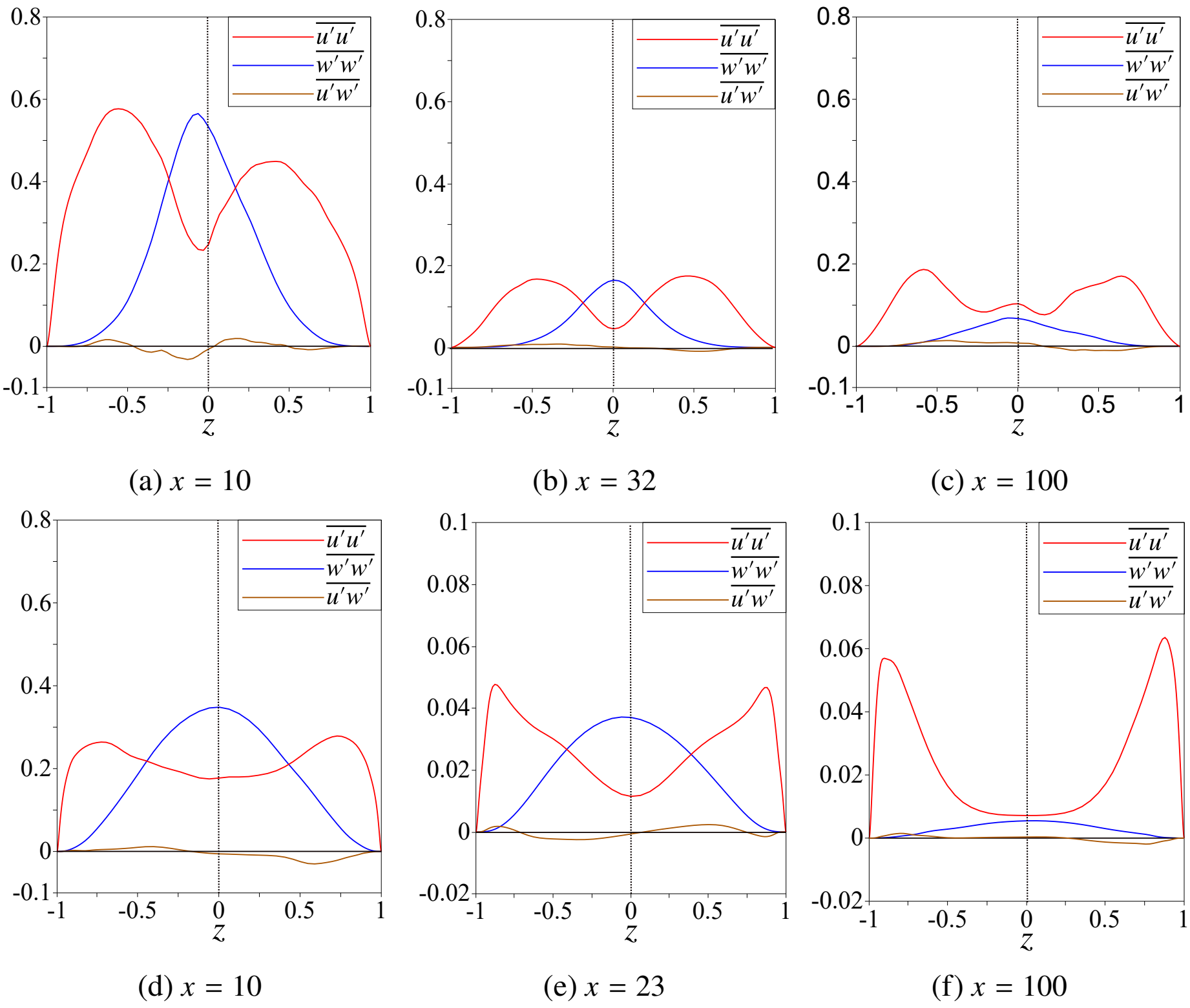

(a) $x$ = 10 (b) $x$ = 32 (c) $x$ = 100

(d) $x$ = 10 (e) $x$ = 23 (f) $x$ = 100

Figure 22. Reynolds stresses between Shercliff walls at $x$ coordinates representative of the Q2D turbulence, transition and buoyancy-driven jet regions for $Ha$ = 325 (a-c) and $Ha$ = 1000 (d-f). Note that all subfigures (a-d) and (e-f) have a different range on the $y-$axis for better readability. Stress components are given in the legend.

### 3.6. *Perfectly insulating, vertical duct with downwards flow: The QW case*

As with the QM flow, in perfectly insulating ducts, when the buoyancy force acts in the opposite direction of the mean flow (point $C$ in figure 5), the flow initially retains the Q2D rolls from the horizontal case, and buoyancy driven jets are formed at the Shercliff walls.

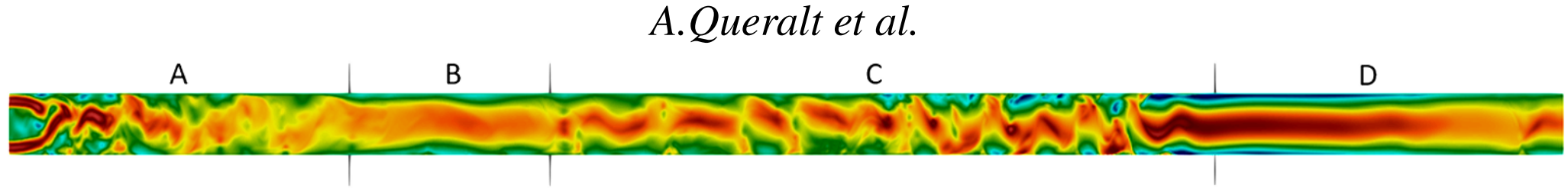

Figure 23. Characteristic $(x, z)$ midplane streamwise velocity segments in a QW flow at $Re = 4000$, $Ha = 1000$ and $Gr = 10^7$.

Unlike the QM flow however, these jets will generate backflow regions, which trigger instabilities in a more violent manner than the QM flow. We call this the QW flow. The use of the letter "$W$" does not imply that the dips in the velocity profile are of the same magnitude as the central peak, and in fact for insulating ducts it is not the case at all. In this case, the magnitude of the average backflow regions is quite small. In section 6, we will see cases where the magnitude of the dips is similar to that of the peak, but without backflow, like in the example from figure 4.

As with the QM flow, the buoyancy-driven side jets are a source of instability. In the QW flow however, the instabilities are more extreme because the change in sign for the velocity results in a much steeper velocity gradient at the Shercliff walls and stronger inflection points. We have observed the same behaviour both with and without vortex promoters. We notice that similar flow structures, including the formation of near-wall jets with strong inflection points, were also observed in earlier studies of MHD downwards flow in the duct with heated Shercliff walls (Belyaev *et al.* 2020). In that work no vortex promoters were used, instead the inlet conditions were generated by the flow past a honeycomb. Like the QM flow, the QW flow is intermittent and caused by the interaction between bulk and boundary layer flow. Once the initial rolls (segment A in figure 23) have been dissipated, the life cycle of the instabilities is as follows:

1. The bulk flow has a positive velocity, while the side jets have a negative velocity. Despite this, the Lorenz force is strong enough to stop inabilities from growing. This state represents segment B in figure 23.
2. As heat is constantly introduced at the Shercliff walls, the Buoyancy force increases, which results in stronger jets, both in magnitude and width. The increase in width of the side jets causes the bulk flow to shrink in the $z-$direction, and by continuity the streamwise velocity has to increase (segment D in figure 23).
3. At a certain (critical) point, the velocity gradients become too large for the Lorenz force to prevent the instabilities from growing. At that point, the side jets break, and once again Q2D vortices populate the duct (segment C).
4. Step 3 destroys the strong velocity gradient, which allows the jets to reform. At the bulk, the Q2D vortices dissipate and the cycle begins anew.

In the QM flow, the jet detachments acted locally, creating Q2D rolls only in their vicinity, and leaving gaps between clusters of rolls. In the QW flow however, when an instability propagates it does so in both directions, towards the inlet by the backflow side jets and towards the outlet by the bulk flow. The result is that whenever an instability appears it affects the flow in the entire duct. This can be seen in figures 24 and 25. We did not conduct simulations with longer ducts to verify if there is a limit to the propagation length of the instabilities. Despite the double propagation, the Q2D rolls are not uniformly distributed along the duct. As with the the QH and QM cases, the time-averaged velocity and Reynolds stresses between Shercliff walls exhibit 2D turbulence like profiles, as shown in figures 26 and 27.

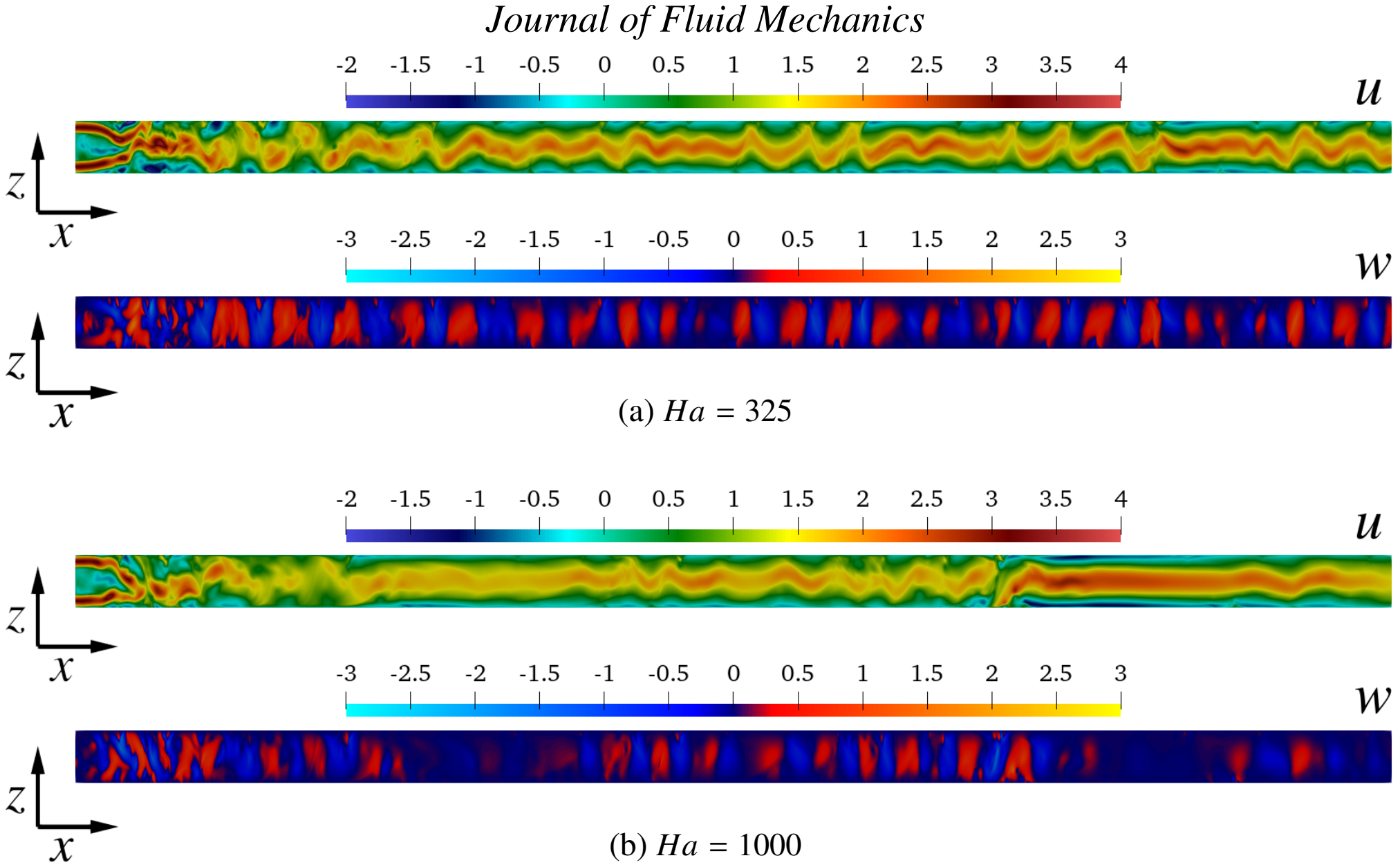

(a) $Ha$ = 325

(b) $Ha$ = 1000

Figure 24. Steamwise $u$ and vertical $w$ velocity components for a downwards flow in a perfectly insulating duct $c_W = 0$ at different $Ha$ and $Re = 4000$. $Gr = 10^7$. Ducts have been rescaled by a factor of 0.5 in the $x$ direction to fit the page.

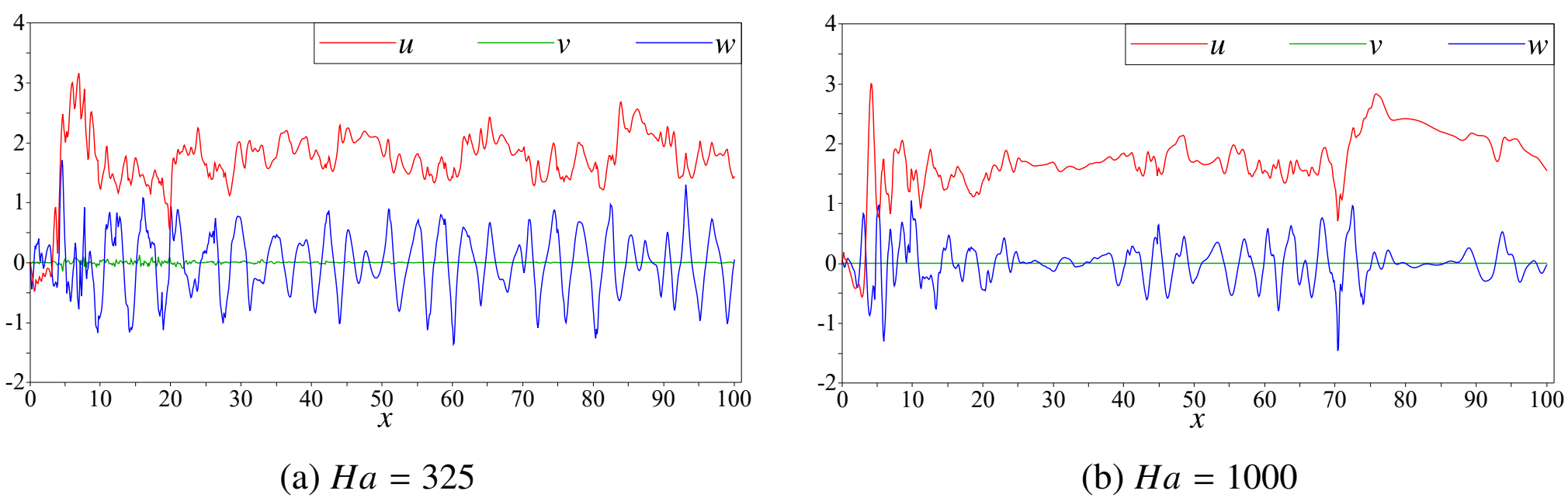

(a) $Ha$ = 325 (b) $Ha$ = 1000

Figure 25. Velocities at the duct centre, downwards flow at $Re = 4000$, $Gr = 10^7$, $c_W = 0$ and different $Ha$. Components are given in the legend.

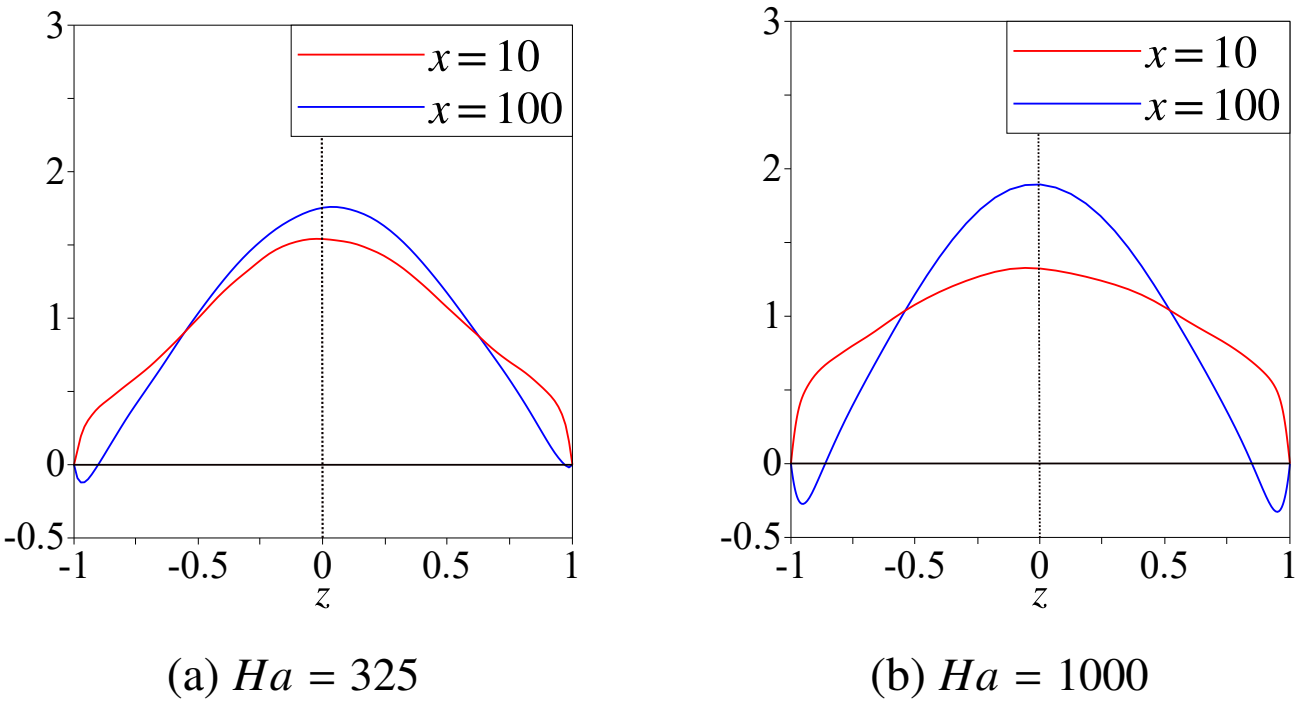

(a) $Ha$ = 325 (b) $Ha$ = 1000

Figure 26. Time-averaged streamwise velocity profiles between Shercliff walls at different duct cross sections. At $Re = 4000$, $Gr = 10^7$, $c_W = 0$ and different values of $Ha$

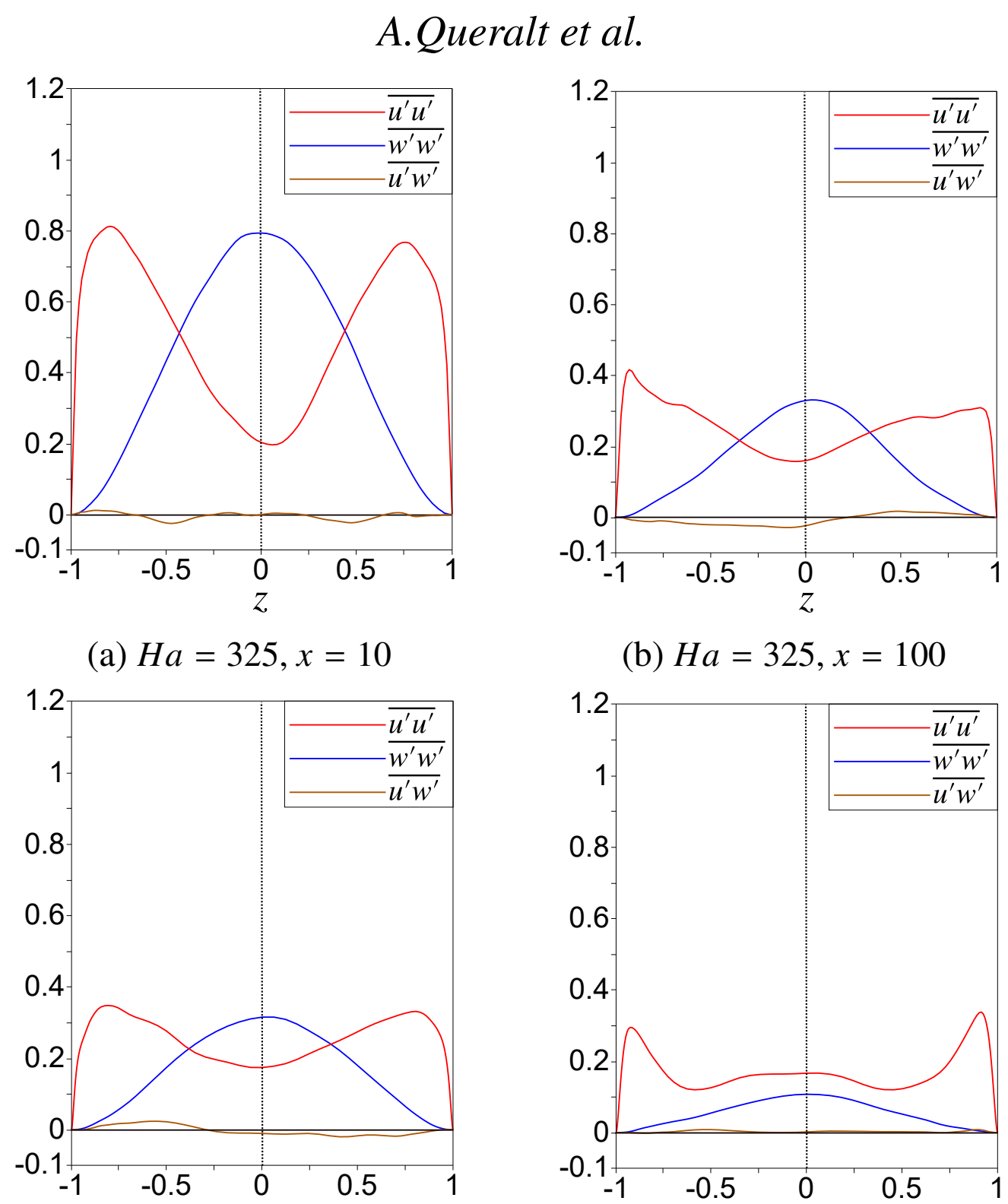

(a) $Ha = 325, x = 10$ (b) $Ha = 325, x = 100$

(c) $Ha = 1000, x = 10$ (d) $Ha = 1000, x = 100$

Figure 27. Time-averaged streamwise velocity profiles between Shercliff walls at different duct cross sections. At $Re = 4000$, $Gr = 10^7$, $c_W = 0$ and different values of $Ha$.

## 4. Momentum and heat transfer properties of the flow regimes

After the description the four flow regimes, we now perform a statistical analysis of each of them (QH, QM, QW and UL). We compare the turbulent kinetic energy $E_{\text{kin}}$ and Nusselt number $Nu$ of each of the flows. We see also a relation between the strength of side jets and $Nu$, which results in some flows with better mixing having a lower $Nu$ due to the lack of side jets.

### 4.1. *Turbulent kinetic Energy*

We calculate the time-averaged turbulent kinetic energy $E_{\text{kin}}(\boldsymbol{x})$ along the duct $x-$direction centreline for all four types of flow, at $Ha = 325,\ 500,\ 750$ and 1000, as shown in figure 28. See definition (3.1) .The effects of the Lorenz and buoyancy forces as well as the duct length are discussed. Then, the volume averaged $\langle E_{\text{kin}}(t)\rangle_V$ is calculated, see figure 29. Its evolution over time shows that the QW flow has the strongest intermittency. Combined with a time-average, we observe cases where a flow with highly conducting walls can have higher kinetic energy than a flow with lower wall conductivity, all else equal.

In cases with electrically insulating walls, the $E_{\text{kin}}(x)$ decays across the duct length as the Q2D structures are dissipated (figure 28a). When heat transfer is added (figures 28c and 28d), $E_{\text{kin}}(x)$ increases once again thanks to the buoyancy-driven instabilities. This increase seems to stabilize after a certain duct length. Energy is lost at the Hartmann walls due to friction, and this process is not compensated by any other mechanism at $Gr = 0$, but when heat transfer is added, the $E_{\mathrm{kin}}(x)$ stabilizes when the rate of loss due to friction at the Hartmann walls is compensated by the thermal energy added at the Shercliff walls.

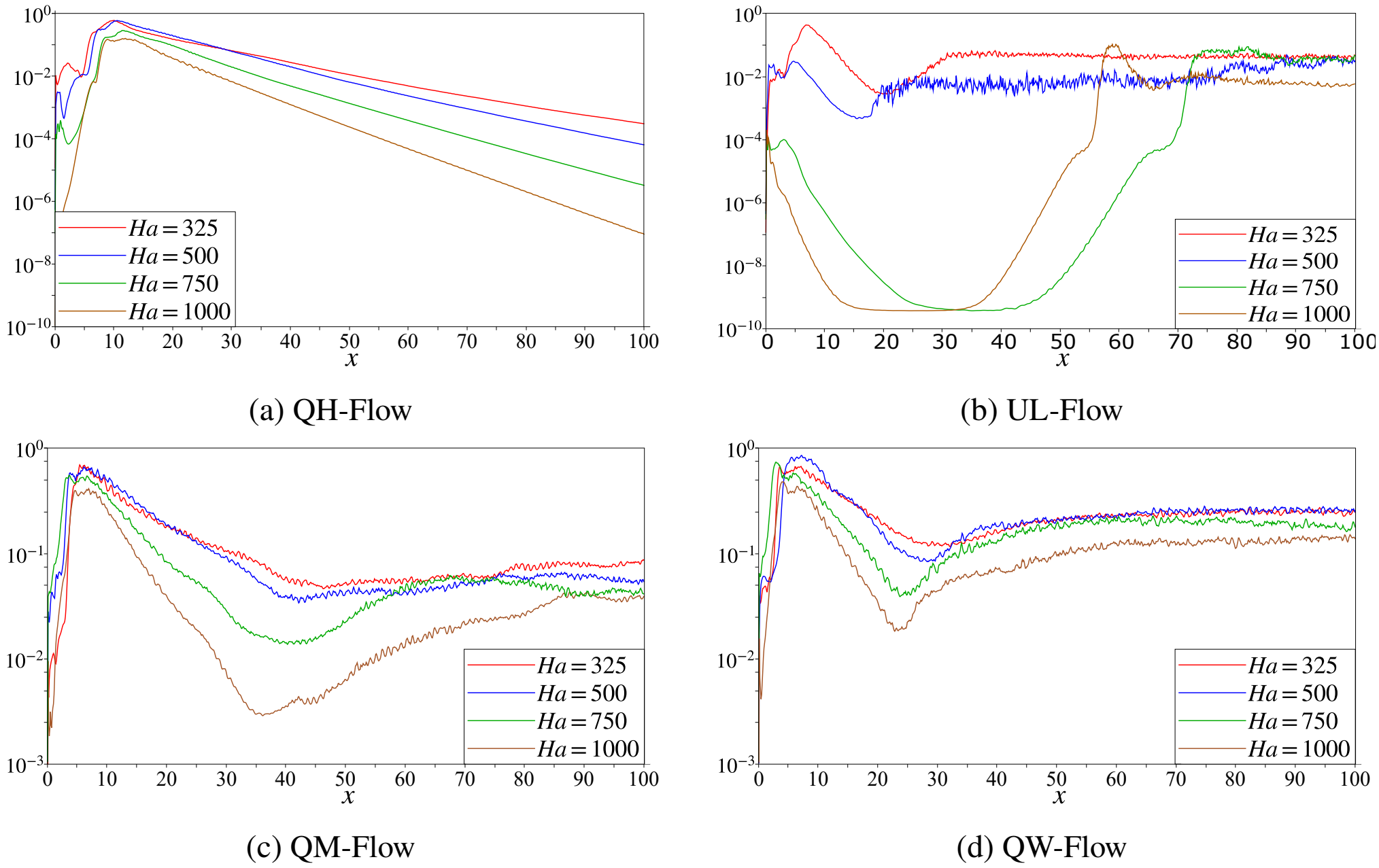

Figure 28. Turbulent kinetic along the duct centerline $x$ at $Re = 4000$ for different $Ha$ (see legends) and the four flow types. This time average is denoted $E_{\mathrm{kin}}(x) = E_{\mathrm{kin}}(x, L_y/2, L_z/2)$. Note that (a) and (b) have a different $y-$axis range than panels (c) and (d).

For highly conducting walls however, the picture changes completely (figure 28b). Firstly, the initial peak of $E_{\mathrm{kin}}(x)$ at the duct inlet is much weaker since Joule dissipation quickly eliminates disturbances. In detachment populated regions the $E_{\mathrm{kin}}(x)$ is roughly constant. For the higher $Ha$, we see a peak of $E_{\mathrm{kin}}(x)$ when the detachments are initially triggered, followed by stabilization later.

From these observations we see that instabilities triggered in ducts with highly conducting walls are not dissipated. This is due to the high vorticity, naturally present in interface between the Shercliff layer and the bulk flow, and the associated inflection points on the velocity profile (see, e.g., Arlt (2018) and references therein).

While the kinetic energy is smaller compared to insulating walls, for very long ducts the volume averaged $\langle E_{\mathrm{kin}}(t)\rangle_V$ can be higher in conducting ducts than in their insulating counterparts. The temporal evolution of the duct-averaged $\langle E_{\mathrm{kin}}(t)\rangle_V$, shown in figure 29 provides an example of a UL flow with higher kinetic energy than a QH flow. In this case, instead of making a longer duct, the effect is due to a change in $Ha$, which as we saw in section 3.4 results in instabilities being triggered earlier. Despite this, both the QH and UL flows have lower $E_{\mathrm{kin}}$ than the QM and QW flows.

Table 4 shows the averages with volume and time and the corresponding standard deviations of the four types of flows at $Ha = 325$ and $Ha = 1000$. The UL flow has not only a higher $E_{\mathrm{kin}}$ than the QH flow at $Ha = 1000$, but it is also the only flow whose kinetic energy increases with $Ha$. The standard deviation $\sigma$ of the volume-averaged $E_{\mathrm{kin}}$ shows that the QW flow has the highest intermittency, and it increases with $Ha$. This is due to the change in velocity direction, instabilities will always be triggered and an increase in $Ha$ only makes the velocity gradients that trigger them greater. The QM flow on the other

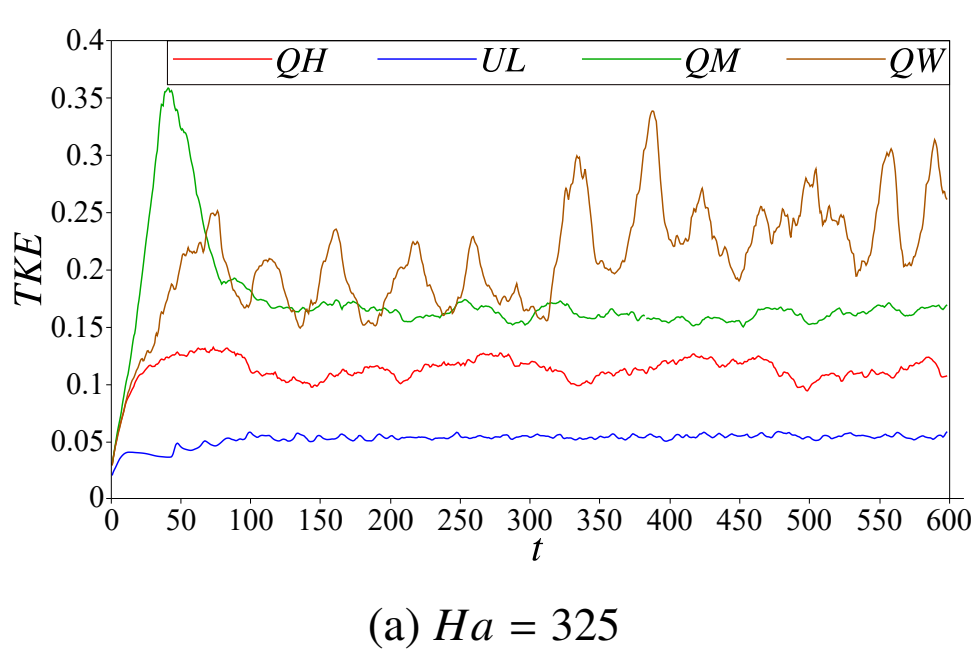

(a) $Ha = 325$

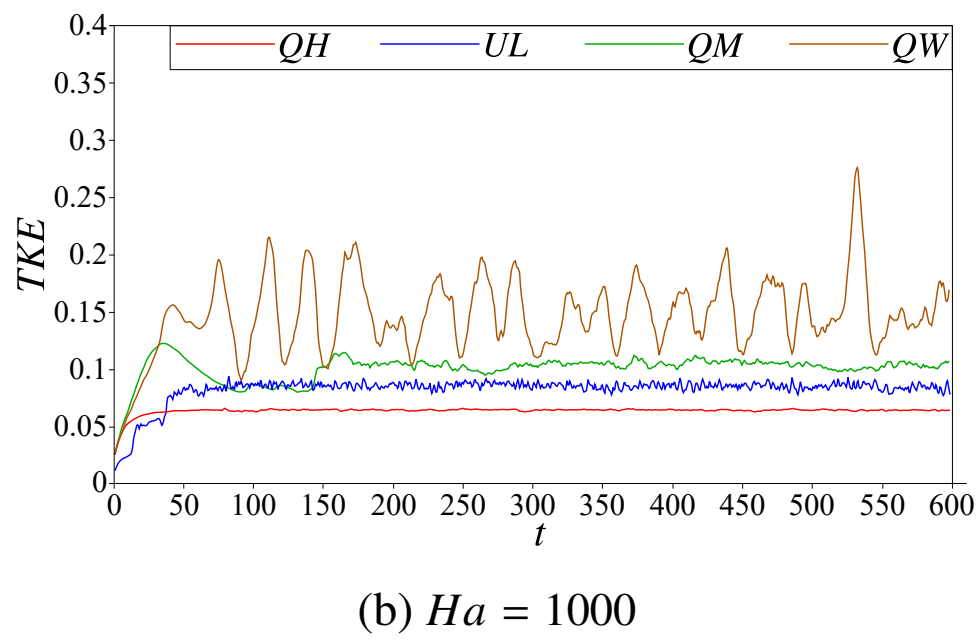

(b) $Ha = 1000$

Figure 29. Temporal evolution of the volume-averaged $\langle E_{\text{kin}}(t)\rangle_V$ for $QH$, $UL$, $QM$ and QW flows, which we denote as TKE on the $y$-axis.

| Flow | $Ha = 325$ $\langle E_{\text{kin}}\rangle_{V,t}$ | $\sigma(\langle E_{\text{kin}}\rangle_V)$ | $Ha = 1000$ $\langle E_{\text{kin}}\rangle_{V,t}$ | $\sigma(\langle E_{\text{kin}}\rangle_V)$ |
|---|---|---|---|---|
| QH | 0.279 | 0.0230 | 0.155 | 0.0041 |
| UL | 0.123 | 0.0051 | 0.190 | 0.0085 |
| QM | 0.335 | 0.0147 | 0.399 | 0.0103 |
| QW | 0.651 | 0.0894 | 0.574 | 0.1030 |

Table 4. Time and volume averaged $E_{\text{kin}}$ for the four flow types at $Ha = 325,\ 1000$ and corresponding standard deviation. The averages are taken over the last 400 convective time units of the simulations

hand sees a decrease in standard deviation, since in that case the side jets are in the same direction as the bulk flow, the increase in $Ha$ contributes to delay the onset of instabilities.

### 4.2. *The Nusselt number*

As a measure of the heat transfer efficiency of each flow type, we calculate the Nusselt number $Nu$ for each of them. The results of analysis, shown further in this section, lead to conclusion that for the parameter space considered, the flows which are better at mixing can also be worse at heat transfer and vice-versa.

We define the Nusselt number assuming a constant, uniform heat flux at the heated segments of the duct (see, e.g., Bejan (2004); Bao *et al.* (2025)):

$$Nu(x,t) = \frac{q_w a}{k(T_w(x,t) - T_b(x,t))}. \tag{4.1}$$

Here $q_w$ is the heat flux to the wall, $k$ the thermal conductivity, $T_w$ the wall temperature averaged in the spanwise $y$-direction $T_w(x) = \langle T_w(x,y)\rangle_y$ and $T_b$ the bulk temperature of the flow

$$T_b(x) = \frac{1}{U_0 L_y L_z} \iint_{L_y\ L_z} uT\,dydz. \tag{4.2}$$

We observed no significant difference when calculating $Nu$ based on one wall or the other, so we present results based only on one wall (top wall for horizontal cases, right wall for upwards flow and left wall for downwards flow). Further we compute integral value of

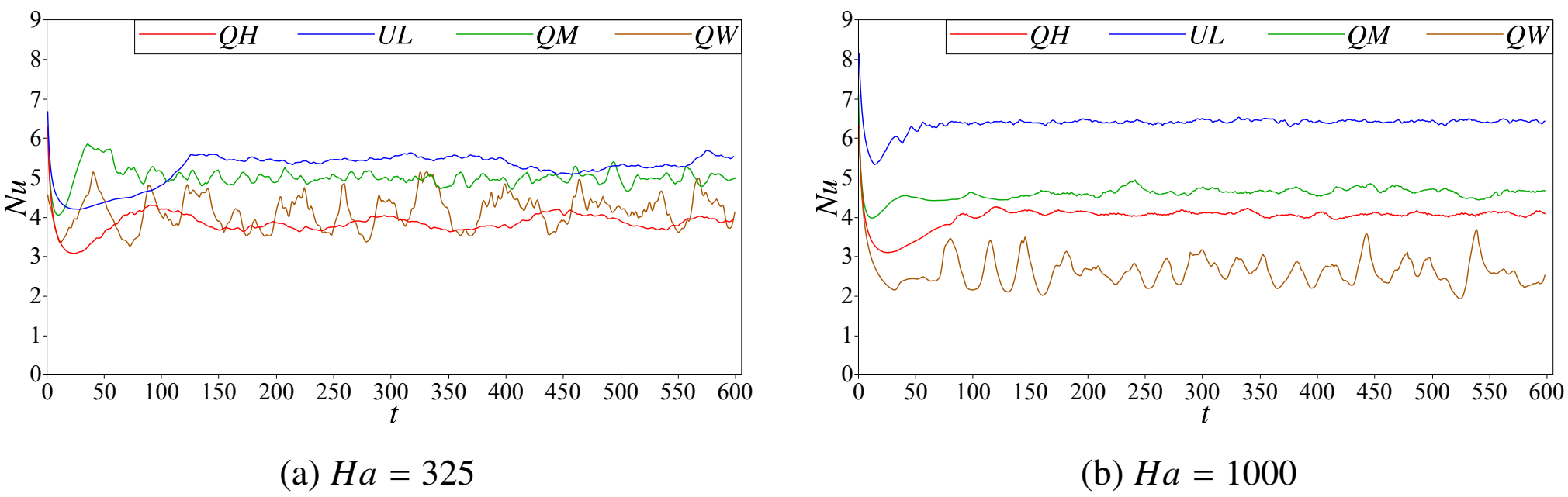

(a) $Ha = 325$ (b) $Ha = 1000$

Figure 30. Temporal evolution of the integral $\langle Nu \rangle_x$ for $QH$, $UL$, $QM$ and QW flows.

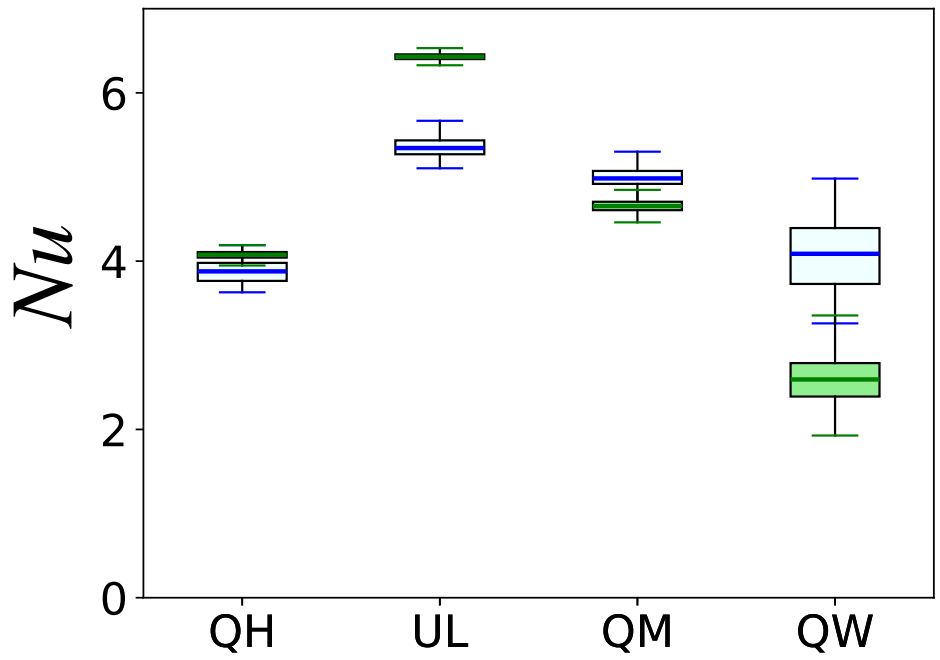

Figure 31. Box plots of the $\langle Nu \rangle_x$ number for $Ha = 325$ (blue) and $Ha = 1000$ (green) for the $QH$, $UL$, $QM$ and QW flows.

$Nu(x, t)$ over the heated section of the duct:

$$\langle Nu \rangle_x = \frac{1}{l - l_0} \int_{l_0}^{l} Nu(x, t) dx \,, \tag{4.3}$$

where $l_0$ marks the $x$ coordinate where the heating begins and $l$ the $x$ coordinate where heating is no longer applied. In our case, $l_0 = 20$ and $l = 32\pi$. Figure 30 shows the temporal evolution of integral $\langle Nu \rangle_x$ for the four flow types at $Ha = 325$ and $Ha = 1000$. In both cases, the UL flow has the highest values. For the other flows, $Nu$ is similar at $Ha = 1000$, but at $Ha = 325$ the QM flow is clearly the second best in terms of heat transfer. Figure 31 compares the values of $\langle Nu \rangle_x$ at both $Ha$. The general trend is that $Nu$ is higher at lower $Ha$, except with the QW flow, where $Nu$ actually decreases.

This leads us to the conclusion that for the flow regimes, that we have considered, there is an inverse relation between $E_{\text{kin}}$ and $Nu$. We suggest a qualitative explanation for this circumstence. We begin by considering a steady case where the vertical velocity $w$ is zero. Therefore, heat is transported in the $z$ direction only by diffusion. The characteristic timescale is

$$\tau_d = a^2/\alpha, \tag{4.4}$$

where $\tau_d$ is the characteristic diffusion timescale. We chose the duct half-height $a$ as a characteristic length because the duct is symmetrically heated and $\alpha$ is the thermal diffusivity $\alpha = \lambda^{-1}$. Heat is advected out from the duct by the streamwise velocity $u$, in a

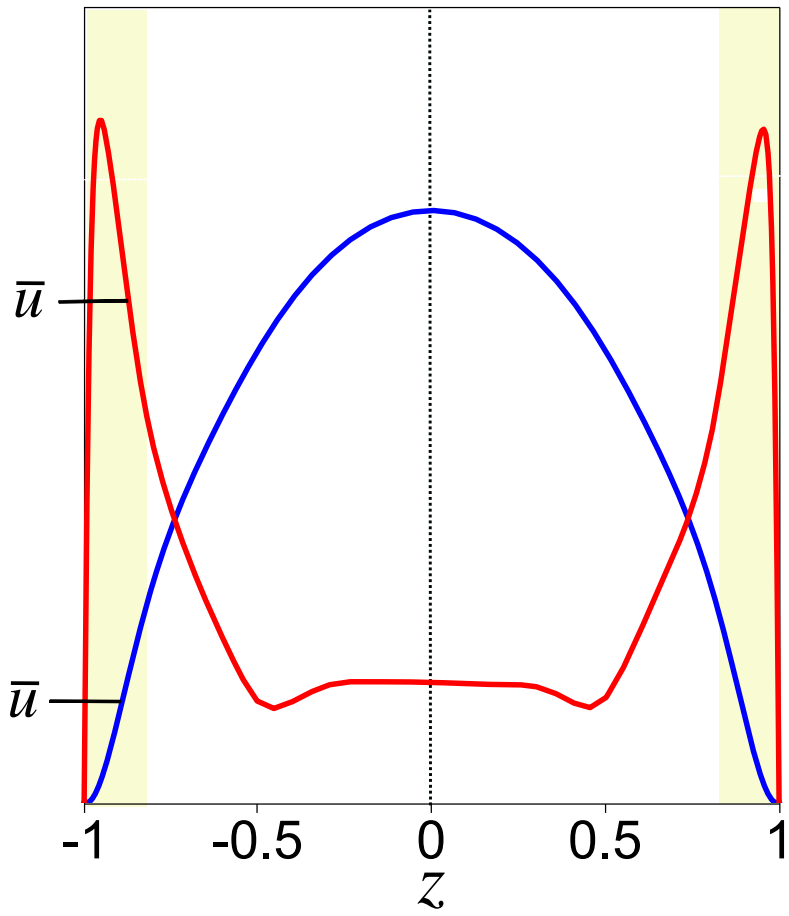

Figure 32. Velocity profiles for electrically insulating (blue) and conducting (red) between Shercliff walls. The vertical velocity is $w \approx 0$, and the yellow shaded area represents the distance heat has diffused after a certain time $\tau_d$.

characteristic timescale

$$\tau_a = L^H / u_c, \tag{4.5}$$

where $\tau_a$ is the characteristic advection timescale. We chose the heated portion of the duct $L^H$ as the characteristic length, and $u_c$ is a characteristic velocity. The distribution of the velocity along $z$ now plays an important role. With strong sidewall jets $\tau_a$ becomes very small, and heat is removed quickly from the duct, even without mixing with the bulk of the flow. When side jets are not present, mixing is required to transport the heated fluid to the bulk, were the mean velocity is larger. This is illustrated in figure 32 which compares the laminar Walker flow and a typical hydrodynamic velocity profile. Calculating the Reynolds number based on the side jet velocity and width can serve as an indicator for the heat transfer efficiency of the flow. Since the fluid is a liquid metal, the diffusive timescale $\tau_d$ is smaller than in most other liquids, which prevents the duct walls from heating up, even when $\tau_d$ is small and the flow is laminar. If we now add the vertical component of the velocity $w$ back and assume that the profiles from figure 32 are averaged profiles rather than a steady state, we notice that $w$ will advect heated portions of the fluid towards the bulk of the flow, which will be beneficial to remove heat from the duct in the insulating case but detrimental in the conducting case. We realize thus that turbulence can in fact be detrimental to heat transfer in cases where strong sidewall jets are present.

## 5. Turbulent kinetic energy production

A phenomenon often associated with two dimensional-turbulence is that of an inverse energy cascade, in which smaller vortices transfer their energy to larger vortices instead of vice-versa (Kraichnan (1967), Batchelor (1969)). In the duct case, energy is transferred to the mean flow by the shear stresses in the boundary layers and the plane jets, the latter of which also existing due to the shear stress (Starr (1968), Votsish & Kolesnikov (1976)). This becomes evident when analysing the turbulent kinetic energy production $P$ which is defined as

$$P = -\overline{u'w'}\frac{\partial \overline{u}}{\partial z} \tag{5.1}$$

We observed the following three production profiles between Shercliff walls:

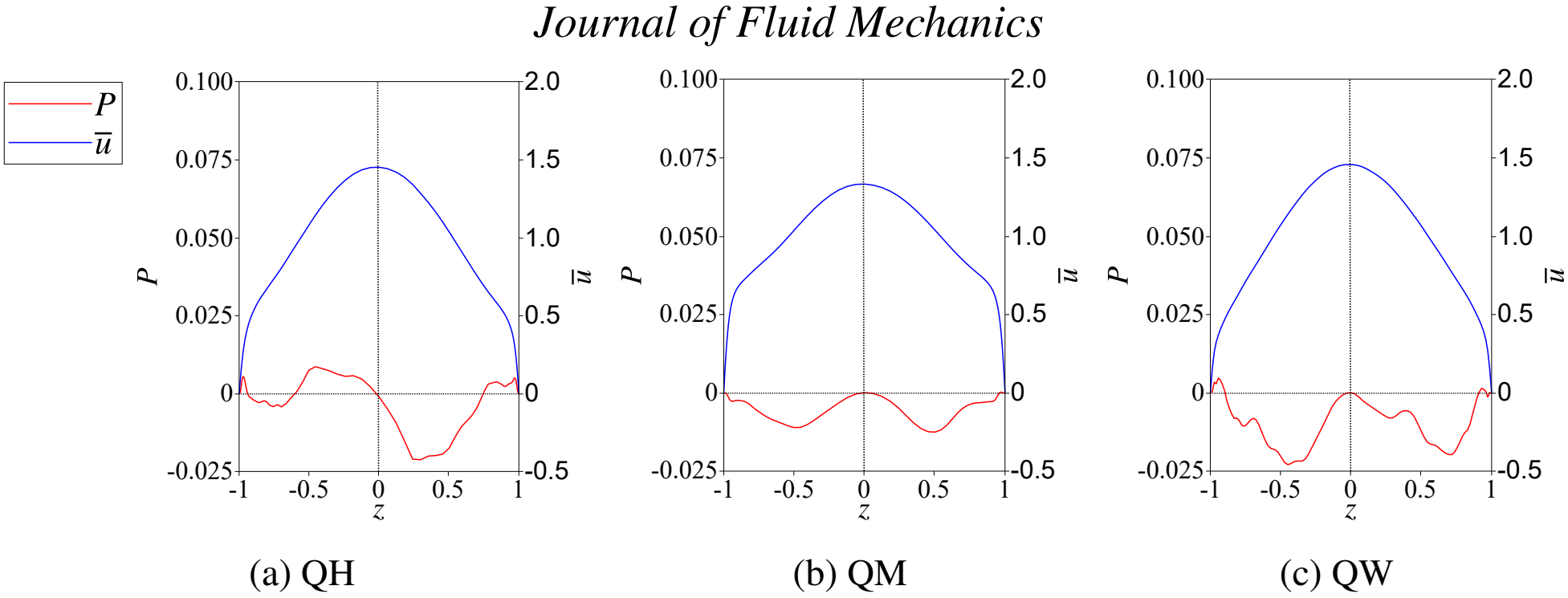

Figure 33. Turbulent kinetic energy production $P$ and time averaged velocity profiles $\overline{u}$ for flows with Q2D structures in the region where they occur $x = 15$, at $Ha = 325$ and $Re = 4000$. In all cases the negative $P$ indicates that the structures are being sustained by an inverse energy cascade.

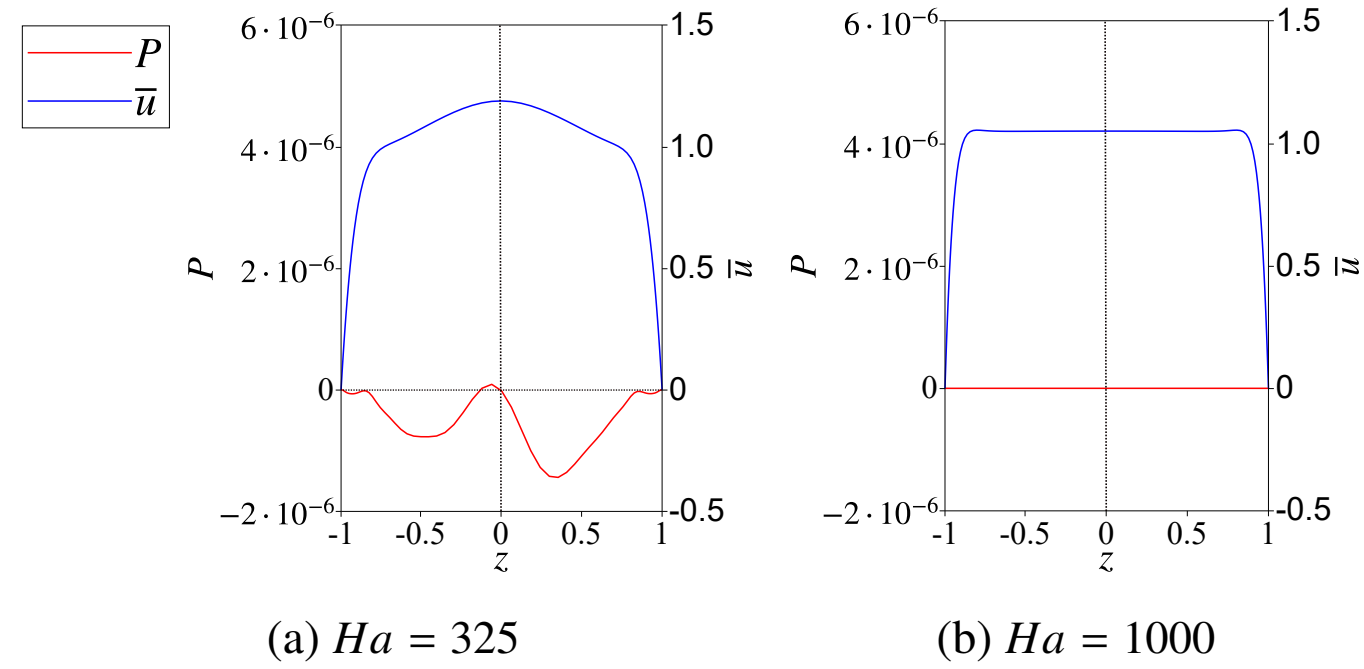

Figure 34. Turbulent kinetic energy production $P$ and time averaged velocity profiles $\overline{u}$ for the QH flow once the Q2D structures have dissipated at $x = 100$, $Re = 4000$ and $Ha = 325$ (a) and $Ha = 1000$ (b). At the lower value of $Ha$, there is still some small residual $P$.

- *Q2D structures:* The Q2D structures present in the QH, QM, and QW all have negative $P$ profiles between Shcerlicff walls, regardless of $Ha$. Figure 33 shows $P$ in the roll-dominated region for each of these three flows, indicating that the inverse energy cascade dominates the flow. Most of the production occurs at $\approx z \pm 0.5$ and the production is symmetric.
- *Hartmann profile:* In the QH flow, once the Q2D structures dissipate, the Hartmann-like profile exhibits no $P$, positive or negative. This is shown in figure 34, where some residual $P$ of the order of $10^{-6}$ is still present in the $Ha = 325$ case. This is because the Q2D structures are not completely dissipated, as was seen in figure 8.
- *Positive side wall jets:* In the QM and UL flows, side jets appeared attached to the Shercliff walls. In the highly conducting wall case at high $Ha$, the jets are initially stable, and in that case there is no $P$ (figure 35a). When the jets become unstable however, a region of negative $P$ appears close to the wall, but the situation is quickly reversed and $P$ becomes positive (figure 35b). The minima and maxima of $P$ are aligned with inflexion points at either side of the jets velocity maxima, while the maxima itself corresponds to $P = 0$. The UL flow has the highest $P$ magnitude, both positive and negative of all the flows we have considered, even when the velocity profile is similar to a non conducting case, such as the QM case, as shown in figures 35c and 35d.

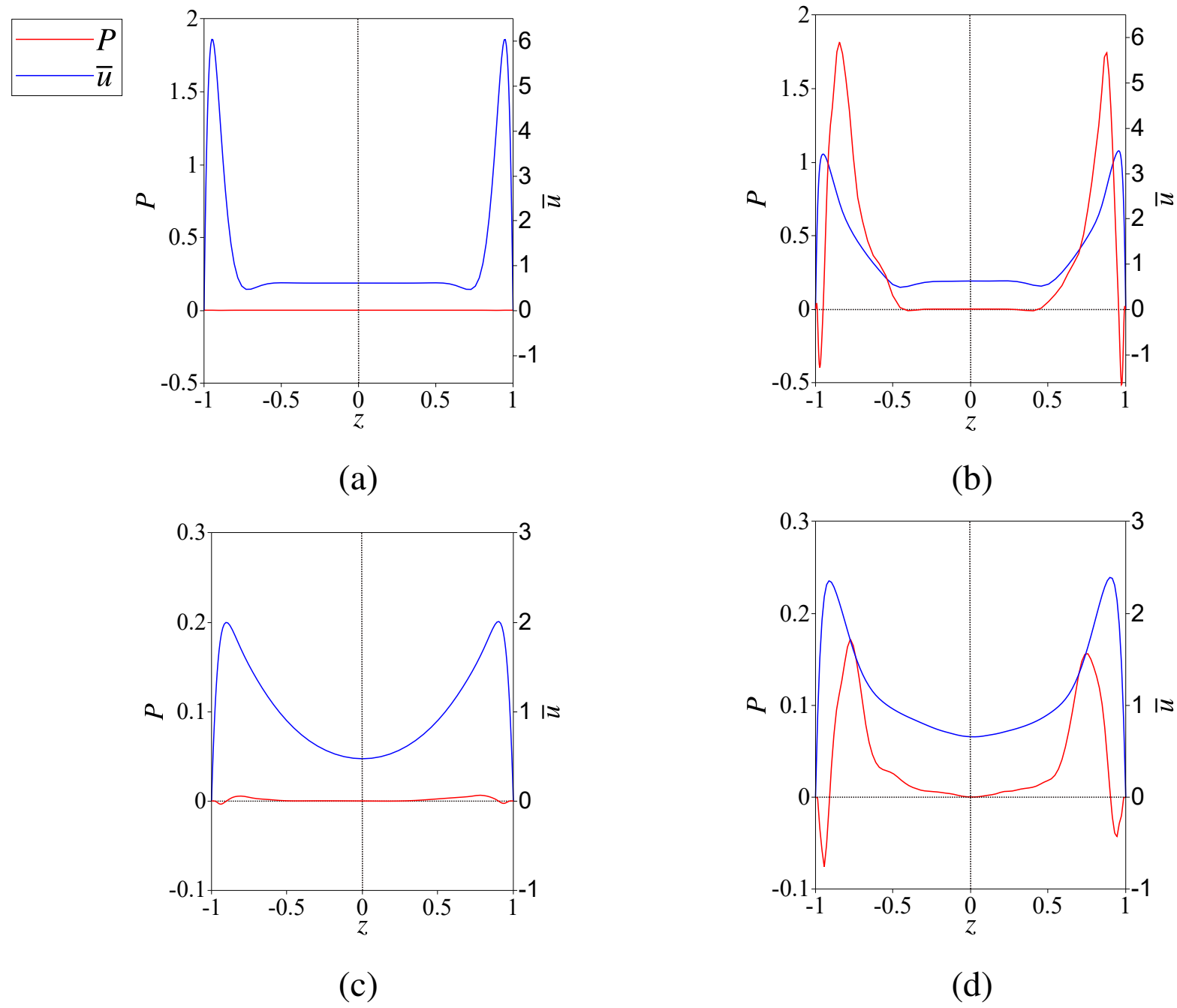

Figure 35. Turbulent kinetic energy production $P$ and time averaged velocity profiles $\overline{u}$ for flows with streamwise side jets. (a) and (b) correspond to the UL flow at $Ha = 1000$, before the jets become unstable at $x = 25$ (a) and at the end of the duct $x = 100$, when the jets are unstable (b). (c) corresponds to the QM flow at $Ha = 325$, $x = 50$. (d) corresponds to the UL flow at $Ha = 325$ and $x = 50$. Despite having a similar velocity profile, $P$ is much larger in the UL case than in the QM case. Note that figures (a,b) have a different scale than figures (c,d).

## 6. Phase diagram of flows at $Re = 4000$ and $Ha = 325$

It has been seen that wall conductivity has a significant effect on the flow, to the point that the effects of buoyancy may become negligible. We now want to determine how large the conductivity has to be for this to occur. To find this transition we ran a series of shorter DNS, with different wall conductance ratios $c_W$, and classify the flow for each simulation. We repeat this procedure at $Gr = 0$, and $Gr \leq 10^7$ both in the upward and downward directions. Appendix A contains a list of all the simulations performed for this purpose. With the results, we construct the "phase diagram", which is shown in figure 36. It is clearly seen that there exist several transition regimes between the four types discussed until now. Due to the high computational costs required, we cannot study them in detail; we will, however give a brief description of each of the ones we have encountered.

We have conducted these simulations at $Re = 4000$ and $Ha = 325$, since these conditions have shown the shortest transients before the flow is fully developed. The simulations were conducted for 150 convective time units. Some parameter combinations had a longer transient before the flow could settle, so we extended those simulations (see appendix A for details).

The flow type will depend on the overall balance between the buoyancy and Lorenz forces acting on the duct. In the four types of flow discussed in sections 3.3 to 3.6, one of the forces always clearly dominated over the other. Now, we have examples where both forces are comparable or, on the contrary, cancel out almost entirely and play no role on the integral characteristics of the flow (other than maintaining a Q2D structure in the case of the Lorenz force).

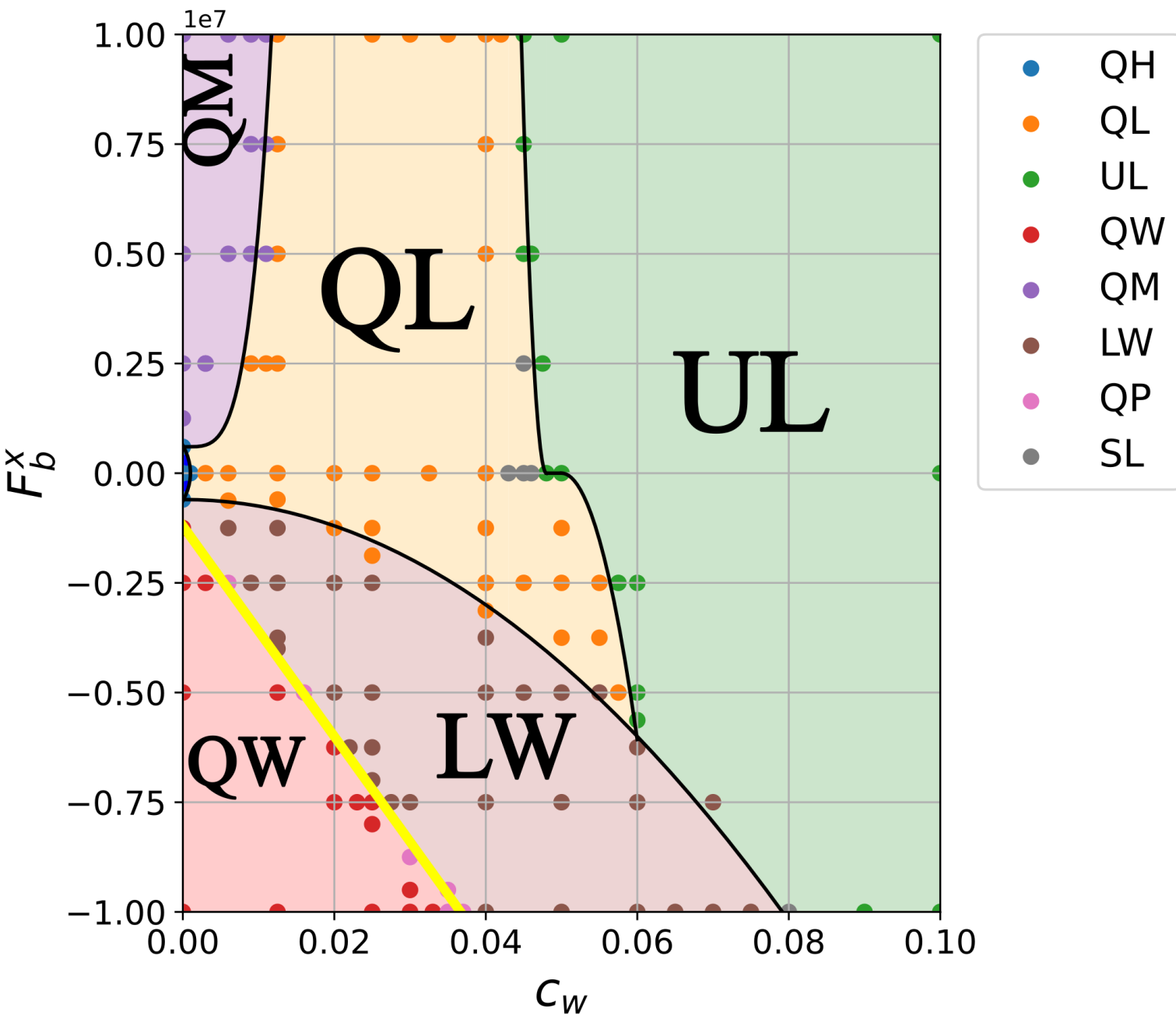

Figure 36. Phase diagram of the different flow types in the $c_W - F_b^x$ plane at $Ha = 325$, $Re = 4000$ and $Pr = 0.02$. For the naming notations see table 2. Details on the runs can be found in the appendix.

Figure 37 shows a broad classification of the conditions in which each type of flow occurs. In what follows, we describe the different flows that occur at constant $F_b^x$ as $c_W$ gradually increases, for $F_b^x \approx 0, F_b^x \gg 0$ and $F_b^x \ll 0$. Of course, it is not physically feasible to change the conductance of the duct dynamically within a real experimental environment, but it makes the description of the phase diagram more intuitive than changing $Gr$ at a fixed $c_W$. We discuss a few specific parameter regimes in more detail.

### 6.1. *The flow at $Gr = 0$ and $c_W \in [0, 0.1]$*

Without the buoyancy force, modifications along the $c_W$ axis represent traversing the plane from figure 5 from point B, where the flow is QH to point E, where the flow is UL. In the parametric study, we saw that the QH flow exists only when the wall conductivity ratio is very close to 0. Between $c_W = 0.001$ and $c_W = 0.003$, side jets begin to form when the Q2D rolls dissipate. We therefore refer to this as a QL flow. The side jets are very weak at first, but grow with $c_W$. Eventually, the Lorenz force is high enough to prevent the Q2D rolls from forming in the first place, and a stable walker flow (SL in our nomenclature) populates the duct. We did not find enough paramaters with this flow to define an area to colour, so we have painted them gray inside of the orange shaded QL flow. As $c_W$ increases further, the jets become stronger and, thanks to the instabilities from he vortex promoters, the jet-detachments are triggered and populate the entire duct. The transition from QL to UL occurs somewhere between $c_W = 0.04$ and $c_W = 0.05$. The simulations conducted at values inside of that range where all SL, but with intermittent instabilities that would die out. We do not rule out that instabilities could be triggered and grow within that range, even if we have not observed them yet. Relation 6.1 illustrates the transitions between regimes

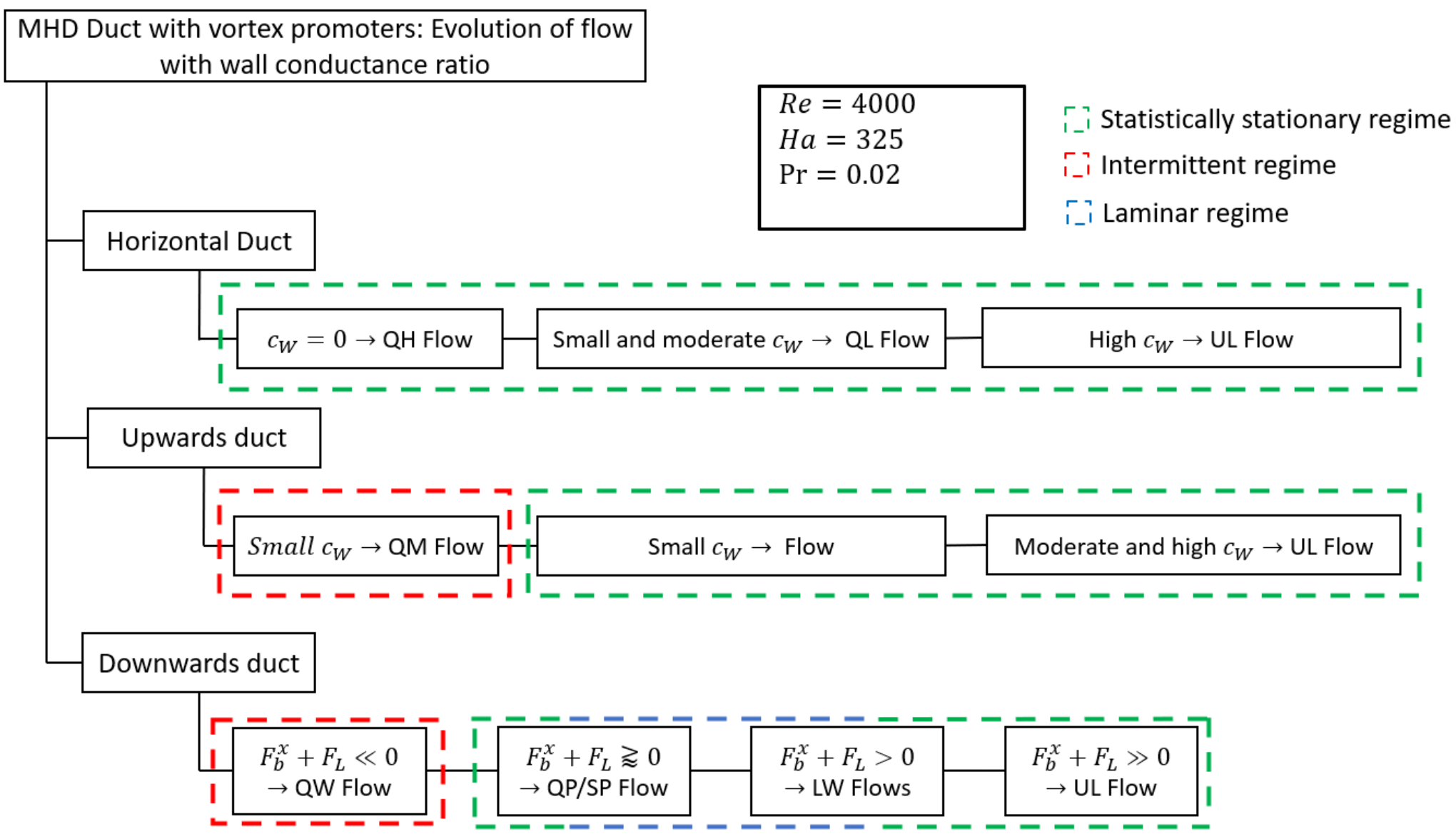

Figure 37. Schematic of the different flow types and causal connections in the $c_W - F_b$ plane at $Ha = 325$, $Re = 4000$ and $Pr = 0.02$.

in these conditions.

$$\left[ \text{QH} \xrightarrow{c_W>0} \text{QL} \xrightarrow{c_W\approx 0.04} \text{SL} \xrightarrow{c_W\approx 0.05} \text{UL} \right]_{Gr=0} \tag{6.1}$$

### 6.2. *The flow at $F_b^x \gg 0$ and $c_W \in [0, 0.1]$*

With the duct in the upwards direction, modifications along the $c_W$ axis represent moving in the upper half of the $F_b^x - c_W$ plane. Outside of the small $Gr$ window where the buoyancy force does not affect the flow, the transitions are similar to the horizontal duct case, except that the QH flow is replaced by the QM, which survives when wall conductivity is added due to the prior existence of side wall jets.

The Lorenz force's effects, namely reduce instabilities and promote Shercliff wall jets, creates stronger, but more stable side jets than those for the insulating duct case. At the highest value of $Gr = 10^7$, the flow transitions to a QL flow at roughly $c_W = 0.0125$. As $c_W$ is increased, once again the initial Q2D rolls are damped and only the unstable side jets of the SL and latter still UL flow. Once again, we have few observations of the SL flow, due to the narrow band where this class of flows is observed. The onset of the UL flow occurs at a lower $c_W$ than for the $Gr = 0$ case. Relation 6.2 illustrates the transitions between regimes for the extreme case of $Gr = 10^7$

$$\left[ \text{QM} \xrightarrow{c_W\approx 0.0125} \text{QL} \xrightarrow{c_W\approx 0.042} \text{UL} \right]_{Gr=10^7 \text{ upwards flow}} \tag{6.2}$$

### 6.3. *The flow at $F_b^x \ll 0$ and $c_W \in [0, 0.1]$*

With the duct in the downwards direction, modifications along the $c_W$ axis represent in the lower half of the $F_b^x - c_W$ plane. This situation has the richest landscape in terms of flow regimes. Unlike the previous case, the jets due to buoyancy and the ones due to the high wall conductance ratio move in opposite directions. We distinguish four possible cases, depending on the balance between the buoyancy and Lorenz jets:

- *Dominant buoyancy backflow jets:* In these conditions, the duct is populated by the QW flow described in section 3.6. The addition of wall conductance causes the backflow regions to reduce in magnitude, and dampens the effects of instabilities. As wall conductivity increases, the transient period before the first instabilities occur gets longer, and, in some cases the flow even appears to remain laminar after 150 convective time units. However, in all the cases we studied, instabilities appeared before 300 convective time units.
- *Balanced sidewall jets:* In this second case, which we have only observed in a narrow band of the parameter space (pink dots around the yellow line in figure 36), the buoyancy and Walker jets balance each other almost completely, and, critically, any imbalance causes no backflow jets. As a result, the flow is determined only by pressure and viscosity, which leads to a Poiseuille-like flow. Figure 38 shows the instantaneous velocity at the midplane for one such case. Figure 39 further shows the velocity profiles at the centreline and between Shercliff walls at the exit of the duct. In these figures, the flow transitions from Q2D rolls to the Poiseuille-like profile, so we call this the QP flow. In other instances there are no rolls and the flow is laminar across the entire duct. We call this second case the SP (Stable Poiseuille, following the nomenclature established in table 2) flow. This flow exits only on a narrow edge between a stable and unstable flow. Increasing wall conductance or reducing buoyancy results in the appearance of stable side jets at the Shercliff walls. On the other hand, reducing the wall conductance or increasing $Gr$ results in the QW flow.

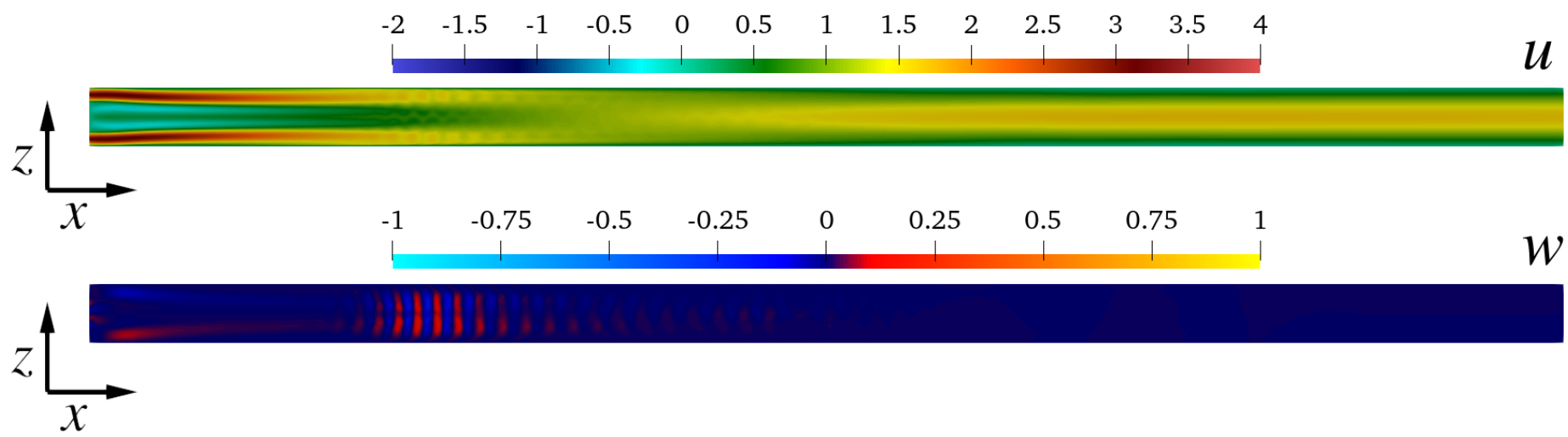

Figure 38. Midplane streamwise $u$ and vertical $w$ velocities at $Re = 4000, Ha = 435, Gr = 10^7$ and $c_W = 0.035$ after 150 convective time units. Note that the range used for the vertical velocity is different than in the previous figures due to the weaker vortices that are formed.

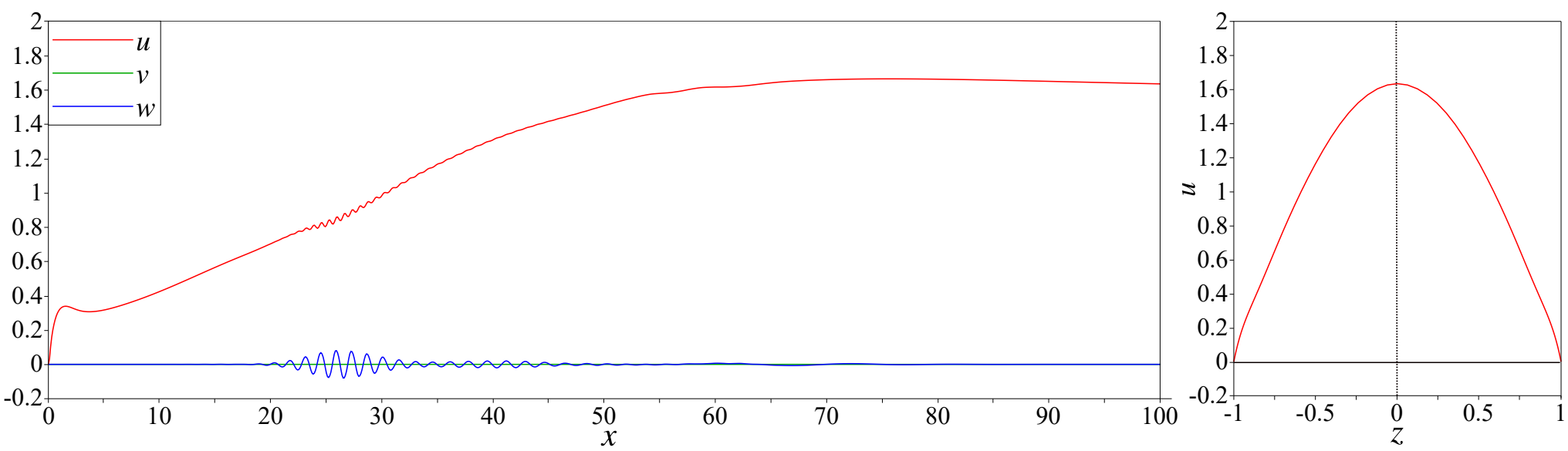

Figure 39. Velocity profiles at the centerline (left) and between Shecliff layers at the duct exit ($x = 100$) (right) for the QP flow at $Re = 4000$, $Ha = 435$, $Gr = 10^7$ and $c_W = 0.035$ after 150 convective time units.

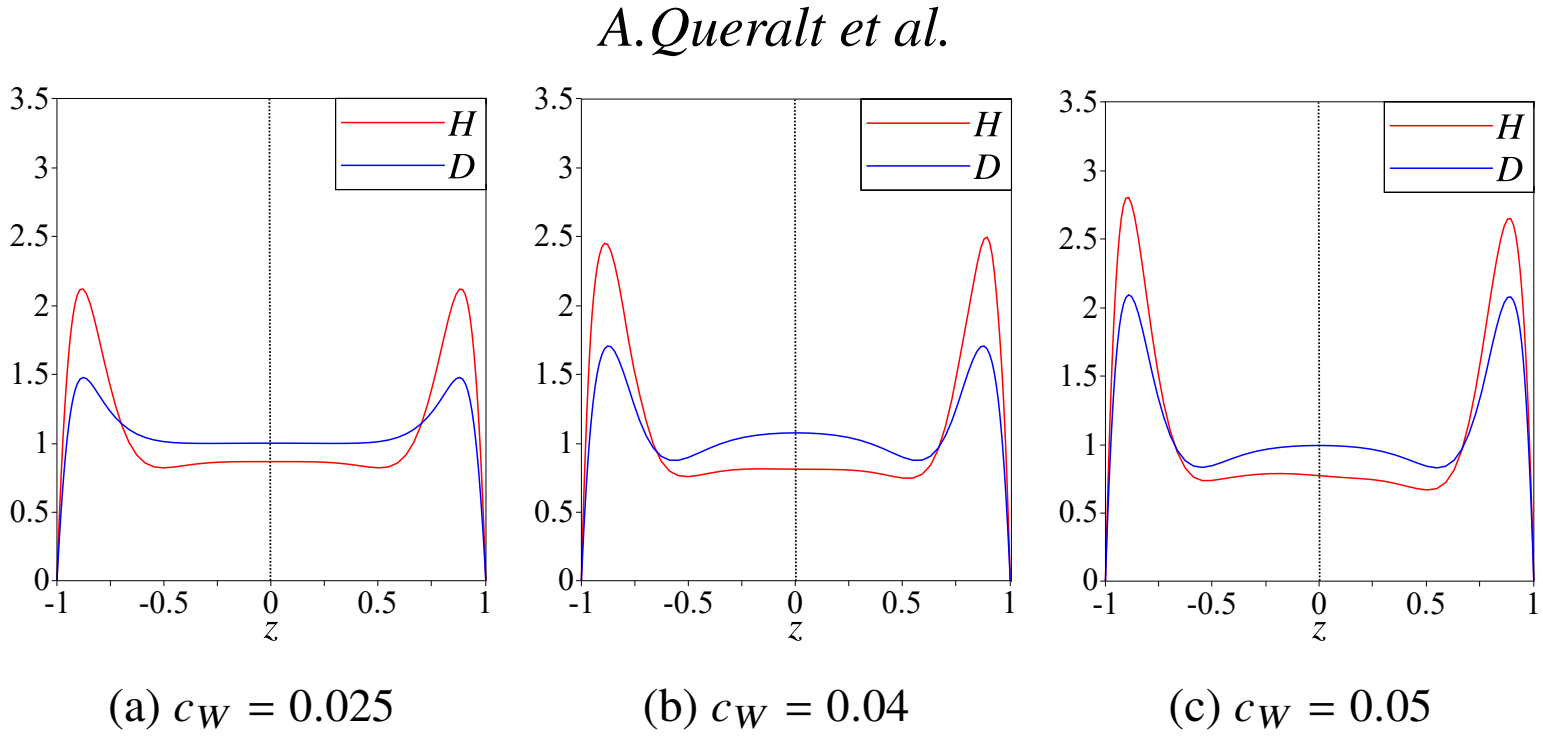

(a) $c_W = 0.025$ (b) $c_W = 0.04$ (c) $c_W = 0.05$

Figure 40. Instantaneous streamwise velocity profile between Shercliff walls for the horizontal (H) duct at $Gr = 0$ and the downward (D) duct at $Gr = 10^7$ both at $Re = 4000$, $Ha = 325$ and three different $c_W$ and $x = 100$

- *Weakly dominanting Walker jets:* As $c_W$ increases, the balance that leads to the Poiseuille-like flow is broken in favour of Walker sidewall jets, which causes the familiar effect of damping the core velocity and forcing the liquid metal to flow in jets close to the Shercliff walls. However, unlike the classic Walker flow, the buoyancy prevents the full formation of side jets and pushes a fraction of the flow back to the core. Figure 40 shows the difference between the side jets formed with and without buoyancy at different $c_W$. When this flow first appears, it still contains Q2D vortices, however these are gradually dissipated until the duct is filled only with side jets. By the time these become unstable, the effects of buoyancy are small and the flow becomes UL. Due to the interplay between both side jets we call this the LW flow.
- *Dominant Walker jets:* At a certain point, the Walker sidewall jets are strong enough that the effects of buoyancy are completely negligible, and the flow takes the UL structure discussed in section 3.4

### 6.4. *Implications for first wall cooling*

The rich landscape of flow regimes provides a great diversity of possible duct configurations to satisfy operational needs in fusion reactors. The UL flow is the best performer in terms of heat transfer, but it is well known that the pressure drop required to maintain flow with conducting walls is too high to be commercially viable, see e.g. Müller & Bühler (2001). In addition, this flow also has the strongest wall velocity gradients, which are also detrimental, see Smolentsev *et al.* (2013). We have seen that side jets occur whenever the duct is insulating and oriented in a vertical or electrically conducting in any orientation. In the former case however, the wall velocity gradients are much smaller, as illustrated in figure 41. In fact, the UL flow not only has the highest velocity gradients, but it also increases drastically with $Ha$.This observation suggests that some loss of $Nu$ can be traded off for a dramatic decrease in near wall velocity gradients and, thus, in total pressure drop $\Delta p$.

## 7. Conclusions and outlook

We have investigated and classified the effect of vortex promoters on the flow structures in liquid metal ducts. Table 5 summarizes the results for the different configurations that we studied jointly. For electrically insulating ducts the effects were important, except when the duct was in a vertical position and the metal was pumped downwards, in which case the MCF remain random. Wall conductivity has a strong effect on these MCF, removing them completely when the wall conductance ratio is high, and potentially reducing intermittency at low or moderate $c_W$, as shown in 6. Research focusing on downward flows with small or moderate wall conductance may yield an optimal set of parameters, such that MCFs loose their random nature and become predictable and controllable.

In addition, the presence of magnetic field and buoyancy forces caused the flow to develop jets at the Shercliff walls. These jets were shown to be supportive for heat transfer, but also introduce strong wall-normal velocity gradients that are undesirable. These jets can be modulated by varying $c_W$, so to an optimal jet-based $Re$ that minimizes pressure drop and wall-normal velocity gradients, while maintaining a fairly high $Nu$.

The length of the duct puts an additional challenge. In highly conducting ducts, the instabilities are generated only after an extended developing length of laminar flow. In the insulating case, when the duct is in a horizontal position, the opposite occurs, the Q2D structures are fastly dissipated, and this is expected to be worse in the fusion environment where $Ha$ is higher.

Based on our results, we can propose two avenues of research that might help tackle these problems. For the conducting wall case, thinner inlet jets can be used as an input. This would be equivalent to placing a larger cylinder at the inlet. If the jets are close to the ones that naturally form at the Shercliff walls they will not be dissipated as quickly by the Lorenz force. At the same time, the jets can be a source of instability if they are not exactly the same as the naturally occurring ones. The idea of using differently sized cylinders, and even changing the position of the cylinder has already been studied, see for example (Hussam & Sheard 2013). However, to the best of our knowledge this has only been done for insulating walls. One might expect that increasing the size of the obstacle would result in a higher pressure drop, however, we note that in the bulk region of the Walker or Hunt flow the velocity is already very low, so increasing the obstacle size might not have as big an impact to pressure drop $\Delta p$ as one might initially expect.

For the insulating case the Q2D structures can be re-triggered by placing obstacles at different positions along the duct. Here, the pressure drop could be affected, so the interval between cylinders should be optimized. Figure 28 hints at the existence of a power or exponential law for the spatial decay of the Q2D structures. This law has to depend (at least) on both $Re$ and $Ha$, and could be related by $R = Re/Ha$ or $N = Ha^2/Re$. With our current data we have not been able to find this law, further exploration of the $Ha - Re$ parameter space is still needed. It must also be noted that this law, if it exists, is only valid in a certain parameter range, the most obvious a priori limitation we can state is that it will work only in the range of $Re$ and $Ha$ where the flow is Q2D.

After having classified the structure formation processes in differently and partly heated ducts for insulating and conducting walls, the control of these configurations to maximize heat transfer would be the next step. These studies have started and will be reported elsewhere.

## 8. Declaration of Interests

The authors report no conflict of interest.

## 9. Acknowledgments

This work is supported by the Deutsche Forschungsegemeinshaft (DFG) with grants KR 4445/5-1 and SCHU 1410/36-1. The authors also acknowledge the Leibniz Supercomputing Centre in Garching, Germany (https://www.lrz.de) for providing computing time on the Supercomputer SuperMUC-NG under the project ‘pn67la’. The flow visualizations and postprocessing were partially done at the University Computing Centre at the TU Ilmenau.

| Duct configuration | | Promoters generate Q2D structures | Performance improvement |
|---|---|---|---|
| Insulating, horizontal | High $Ha$ | Yes | Yes |
| | Low $Ha$ | | |
| Insulating, upward | High $Ha$ | Yes | Yes |
| | Low $Ha$ | | |
| Insulating downward | High $Ha$ | No, Q2D structures are naturally present | NA |
| | Low $Ha$ | | |
| Highly conducting, all orientations | High $Ha$ | No, detachments appear downstream naturally | NA |
| | Low $Ha$ | Yes | Yes |

Table 5. Summary of the effect of the vortex promoters on the flow in terms of both the generation of Q2D structures and heat transfer performance, for each duct configuration.

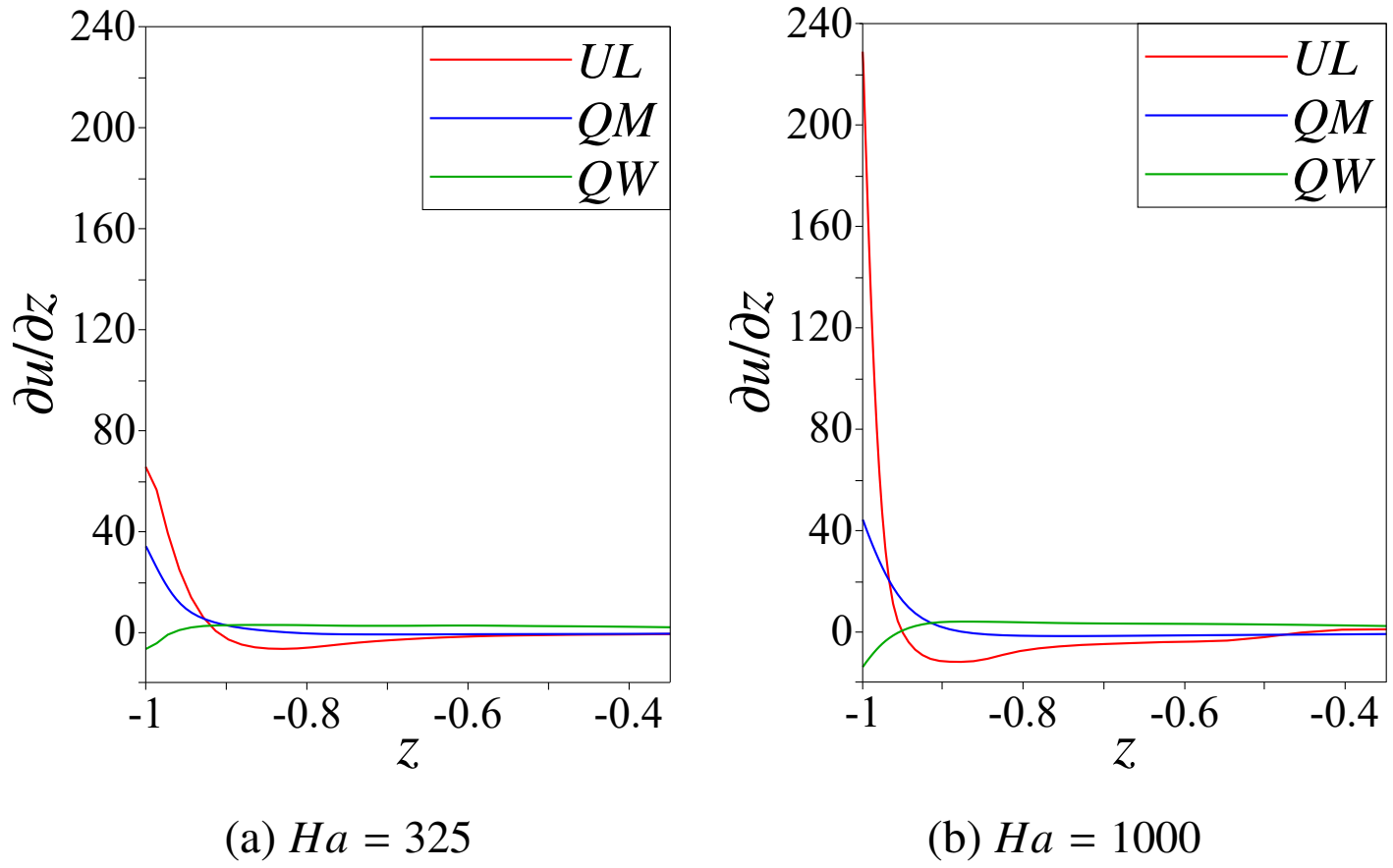

(a) $Ha$ = 325 (b) $Ha$ = 1000

Figure 41. Near wall streamwise velocity gradients at $x$ = 100 for the flows with side jets. Note that due to the backflow jets, the QW gradient is negative.

## Appendix A. Simulation Parameters for the Phase Diagram

Tables 6, 7 and 8 show the physical parameters, simulation duration and resulting flow type for each of the simulations performed for section 6. Horizontal lines indicate a change in $Gr$, and for each $Gr$ the wall conductance ratio is in ascending order. For the Horizontal and Upwards flow, the simulation length was always 150 convective time units (one and a half turnover lengths of the duct). For some of the Downwards simulations, particularly those close to the limit between the QW and the QP flows, some simulations where extended to 300 convective time units or 3 turnover times of the duct.

| $Gr$ | $c_W$ | Flow Direction | Simulated time | Flow Type |
|---|---|---|---|---|
| 0.0 | 0.0000 | NA | 600 | QH |
| 0.0 | 0.0010 | NA | 150 | QH |
| 0.0 | 0.0030 | NA | 150 | QL |
| 0.0 | 0.0060 | NA | 150 | QL |
| 0.0 | 0.0125 | NA | 150 | QL |
| 0.0 | 0.0200 | NA | 150 | QL |
| 0.0 | 0.0250 | NA | 100 | QL |
| 0.0 | 0.0325 | NA | 150 | QL |
| 0.0 | 0.0400 | NA | 150 | QL |
| 0.0 | 0.0430 | NA | 150 | SL |
| 0.0 | 0.0450 | NA | 150 | SL |
| 0.0 | 0.0460 | NA | 150 | SL |
| 0.0 | 0.0480 | NA | 150 | UL |
| 0.0 | 0.0500 | NA | 150 | UL |
| 0.0 | 0.1000 | NA | 600 | UL |

Table 6. Parameters and results of the simulations conducted for section 6. All simulations performed at $Re = 4000$, $Ha = 325$ and $Pr = 0.02$, without heat transfer

| $Gr$ | $c_W$ | Flow Direction | Simulated time | Flow Type |
|---|---|---|---|---|
| $10^7$ | 0.0000 | upwards | 150 | QM |
| $10^7$ | 0.0060 | upwards | 150 | QM |
| $10^7$ | 0.0090 | upwards | 150 | QM |
| $10^7$ | 0.0011 | upwards | 150 | QM |
| $10^7$ | 0.0125 | upwards | 150 | QL |
| $10^7$ | 0.0250 | upwards | 150 | QL |
| $10^7$ | 0.0300 | upwards | 150 | QL |
| $10^7$ | 0.0350 | upwards | 150 | QL |
| $10^7$ | 0.0400 | upwards | 150 | QL |

| $10^7$ | 0.0420 | upwards | 150 | QL |
|---|---|---|---|---|
| $10^7$ | 0.0450 | upwards | 150 | UL |
| $10^7$ | 0.0500 | upwards | 150 | UL |
| $10^7$ | 0.1000 | upwards | 600 | UL |
| $7.5 \cdot 10^6$ | 0.0090 | upwards | 150 | QM |
| $7.5 \cdot 10^6$ | 0.0110 | upwards | 150 | QM |
| $7.5 \cdot 10^6$ | 0.0125 | upwards | 150 | QL |
| $7.5 \cdot 10^6$ | 0.0400 | upwards | 150 | QL |
| $7.5 \cdot 10^6$ | 0.0450 | upwards | 150 | UL |
| $5 \cdot 10^6$ | 0.0000 | upwards | 150 | QM |
| $5 \cdot 10^6$ | 0.0060 | upwards | 150 | QM |
| $5 \cdot 10^6$ | 0.0090 | upwards | 150 | QM |
| $5 \cdot 10^6$ | 0.0011 | upwards | 150 | QM |
| $5 \cdot 10^6$ | 0.0125 | upwards | 150 | QL |
| $5 \cdot 10^6$ | 0.0400 | upwards | 150 | QL |
| $5 \cdot 10^6$ | 0.0450 | upwards | 150 | UL |
| $5 \cdot 10^6$ | 0.0460 | upwards | 150 | UL |
| $2.5 \cdot 10^6$ | 0.0000 | upwards | 150 | QM |
| $2.5 \cdot 10^6$ | 0.0030 | upwards | 150 | QM |
| $2.5 \cdot 10^6$ | 0.0090 | upwards | 150 | QL |
| $2.5 \cdot 10^6$ | 0.0110 | upwards | 150 | QL |
| $2.5 \cdot 10^6$ | 0.0125 | upwards | 150 | QL |
| $2.5 \cdot 10^6$ | 0.0450 | upwards | 150 | SL |
| $2.5 \cdot 10^6$ | 0.0475 | upwards | 150 | UL |
| $1.25 \cdot 10^6$ | 0.0000 | upwards | 150 | QM |
| $6 \cdot 10^5$ | 0.0000 | upwards | 150 | QH |

Table 7. Parameters and results of the simulations conducted for section 6. All simulations performed at $Re = 4000$, $Ha = 325$ and $Pr = 0.02$, with heat transfer and flow directed upwards

| $Gr$ | $c_W$ | Flow Direction | Simulated time | Flow Type |
|---|---|---|---|---|
| $10^7$ | 0.0000 | downwards | 150 | QW |
| $10^7$ | 0.0125 | downwards | 150 | QW |
| $10^7$ | 0.0250 | downwards | 150 | QW |
| $10^7$ | 0.0300 | downwards | 300 | QW |
| $10^7$ | 0.0330 | downwards | 300 | QW |
| $10^7$ | 0.0350 | downwards | 150 | QP |
| $10^7$ | 0.0370 | downwards | 150 | QP |
| $10^7$ | 0.0400 | downwards | 150 | LW |
| $10^7$ | 0.0500 | downwards | 150 | LW |
| $10^7$ | 0.0600 | downwards | 150 | LW |
| $10^7$ | 0.0650 | downwards | 150 | LW |
| $10^7$ | 0.0700 | downwards | 150 | LW |
| $10^7$ | 0.0750 | downwards | 150 | LW |
| $10^7$ | 0.0800 | downwards | 150 | SL |
| $10^7$ | 0.0900 | downwards | 150 | UL |
| $10^7$ | 0.1000 | downwards | 600 | UL |
| $9.5 \cdot 10^6$ | 0.0300 | downwards | 150 | QW |
| $9.5 \cdot 10^6$ | 0.0350 | downwards | 150 | QP |
| $8.75 \cdot 10^6$ | 0.0300 | downwards | 150 | QP |
| $8 \cdot 10^6$ | 0.0250 | downwards | 300 | QW |
| $7.5 \cdot 10^6$ | 0.0200 | downwards | 150 | QW |
| $7.5 \cdot 10^6$ | 0.0230 | downwards | 150 | QW |
| $7.5 \cdot 10^6$ | 0.0250 | downwards | 300 | QW |
| $7.5 \cdot 10^6$ | 0.0275 | downwards | 150 | LW |
| $7.5 \cdot 10^6$ | 0.0300 | downwards | 150 | LW |
| $7.5 \cdot 10^6$ | 0.0400 | downwards | 300 | LW |
| $7.5 \cdot 10^6$ | 0.0500 | downwards | 150 | LW |
| $7.5 \cdot 10^6$ | 0.0600 | downwards | 150 | LW |
| $7.5 \cdot 10^6$ | 0.0700 | downwards | 150 | LW |
| $7 \cdot 10^6$ | 0.0250 | downwards | 150 | LW |
| $6.25 \cdot 10^6$ | 0.0200 | downwards | 300 | QW |
| $6.25 \cdot 10^6$ | 0.0220 | downwards | 150 | LW |
| $6.25 \cdot 10^6$ | 0.0250 | downwards | 150 | LW |
| $6.25 \cdot 10^6$ | 0.0600 | downwards | 150 | LW |
| $5 \cdot 10^6$ | 0.0000 | downwards | 200 | QW |
| $5 \cdot 10^6$ | 0.0125 | downwards | 200 | QW |
| $5 \cdot 10^6$ | 0.0160 | downwards | 150 | QP |
| $5 \cdot 10^6$ | 0.0200 | downwards | 150 | LW |
| $5 \cdot 10^6$ | 0.0250 | downwards | 150 | LW |
| $5 \cdot 10^6$ | 0.0400 | downwards | 150 | LW |
| $5 \cdot 10^6$ | 0.0450 | downwards | 150 | LW |
| $5 \cdot 10^6$ | 0.0500 | downwards | 150 | LW |
| $5 \cdot 10^6$ | 0.0550 | downwards | 150 | LW |
| $5 \cdot 10^6$ | 0.0575 | downwards | 150 | QL |
| $5 \cdot 10^6$ | 0.0600 | downwards | 150 | UL |
| $4 \cdot 10^6$ | 0.0125 | downwards | 150 | LW |
| $3.75 \cdot 10^6$ | 0.0125 | downwards | 150 | LW |
| $3.75 \cdot 10^6$ | 0.0400 | downwards | 150 | LW |
| $3.75 \cdot 10^6$ | 0.0500 | downwards | 150 | QL |
| $3.75 \cdot 10^6$ | 0.0550 | downwards | 150 | QL |
| $3.125 \cdot 10^6$ | 0.0400 | downwards | 150 | QL |
| $2.5 \cdot 10^6$ | 0.0000 | downwards | 150 | QW |
| $2.5 \cdot 10^6$ | 0.0030 | downwards | 150 | QW |
| $2.5 \cdot 10^6$ | 0.0060 | downwards | 300 | QP |
| $2.5 \cdot 10^6$ | 0.0090 | downwards | 150 | LW |
| $2.5 \cdot 10^6$ | 0.0125 | downwards | 150 | LW |
| $2.5 \cdot 10^6$ | 0.0200 | downwards | 150 | LW |
| $2.5 \cdot 10^6$ | 0.0250 | downwards | 150 | LW |
| $2.5 \cdot 10^6$ | 0.0400 | downwards | 150 | QL |
| $2.5 \cdot 10^6$ | 0.0450 | downwards | 150 | QL |
| $2.5 \cdot 10^6$ | 0.0500 | downwards | 150 | QL |
| $2.5 \cdot 10^6$ | 0.0550 | downwards | 150 | QL |
| $2.5 \cdot 10^6$ | 0.0575 | downwards | 150 | UL |
| $2.5 \cdot 10^6$ | 0.0600 | downwards | 150 | QL |
| $1.88 \cdot 10^6$ | 0.0250 | downwards | 150 | QL |
| $1.25 \cdot 10^6$ | 0.0000 | downwards | 150 | QW |
| $1.25 \cdot 10^6$ | 0.0060 | downwards | 150 | LW |
| $1.25 \cdot 10^6$ | 0.0125 | downwards | 150 | LW |
| $1.25 \cdot 10^6$ | 0.0200 | downwards | 150 | QL |
| $1.25 \cdot 10^6$ | 0.0250 | downwards | 150 | QL |
| $1.25 \cdot 10^6$ | 0.0400 | downwards | 150 | QL |
| $1.25 \cdot 10^6$ | 0.0500 | downwards | 150 | QL |
| $6.25 \cdot 10^5$ | 0.0060 | downwards | 150 | QL |
| $6 \cdot 10^5$ | 0.0000 | downwards | 150 | QH |
| $6 \cdot 10^5$ | 0.0125 | downwards | 150 | QL |

Table 8. Parameters and results of the simulations conducted for section 6. All simulations performed at $Re = 4000$, $Ha = 325$ and $Pr = 0.02$, with heat transfer and flow directed downwards